\documentclass[aoas,MSNbibl,nameyear,dvips]{arximspdf}
\usepackage{graphicx}
\doi{10.1214/11-AOAS478}
\volume{5}
\issue{4}
\pubyear{2011}
\firstpage{2519}
\lastpage{2548}

\makeatletter
\newcommand{\eqref}[1]{(\ref{#1})}

\makeatother

\begin{document}
\begin{frontmatter}

\title{Covariance approximation for large multivariate spatial data
sets with an application to multiple~climate model errors\thanksref{T1}}
\runtitle{Covariance approximation}

\begin{aug}
\author{\fnms{Huiyan} \snm{Sang}\corref{}\ead[label=e1]{huiyan@stat.tamu.edu}\thanksref{t2}},
\author{\fnms{Mikyoung} \snm{Jun}\ead[label=e2]{mjun@stat.tamu.edu}\thanksref{t3}}
\and
\author{\fnms{Jianhua Z.} \snm{Huang}\ead[label=e3]{jianhua@stat.tamu.edu}\thanksref{t2,t4}}
\thankstext{T1}{This publication is based in part on work supported by Award No. KUS-C1-016-04, made by
King Abdullah University of Science and Technology (KAUST).}
\thankstext{t2}{Supported in part by NSF Grant DMS-10-07618.}
\thankstext{t3}{Supported in part by NSF Grant DMS-09-06532.}
\thankstext{t4}{Supported in part by NSF Grant DMS-09-07170 and the NCI
Grant CA57030.}
\runauthor{H. Sang, M. Jun and J. Z. Huang}
\affiliation{Texas A\&M University}
\address{Department of Statistics\\
Texas A\&M University\\
College Station, Texas\\
USA\\
\printead{e1}\\
\phantom{E-mail: }\printead*{e2}\\
\phantom{E-mail: }\printead*{e3}}

\end{aug}

\received{\smonth{11} \syear{2010}}
\revised{\smonth{4} \syear{2011}}

%
\begin{abstract}
This paper investigates the cross-correlations across multiple climate
model errors. We build
a Bayesian hierarchical model that accounts for the spatial dependence
of individual models as well as cross-covariances across different
climate models. Our method allows for a nonseparable and nonstationary
cross-covariance structure. We also present a covariance
approximation approach to facilitate the computation in the modeling
and analysis of very large multivariate spatial data sets. The covariance
approximation consists of two parts: a reduced-rank part to capture
the large-scale spatial dependence, and a sparse covariance matrix
to correct the small-scale dependence error induced by the
reduced rank approximation. We pay special attention to the case that
the second part of the approximation has a block-diagonal structure.
Simulation results of model fitting and prediction show substantial
improvement of the proposed approximation over the predictive
process approximation and the independent blocks analysis. We then
apply our computational approach to the joint statistical modeling
of multiple climate model errors.
\end{abstract}

%

\end{frontmatter}
%

\section{Introduction}\label{sec1}

This paper addresses the problem of combining
multiple climate model outputs while accounting for dependence across
different models as well as spatial dependence within each individual
model. To study the impact of human activity on climate change, the
Intergovernmental Panel on Climate Change (IPCC) is coordinating
efforts worldwide to develop coupled atmosphere-ocean general
circulation models (AOGCMs). Various organizations around the world
are developing state-of-the-art numerical models and currently 20$+$
climate models are available. A growing body of literature also exists
that studies multiple climate model outputs
[e.g., \citet{tebaldietal05};
\citet{furreretal07};
\citeauthor{junetal081} (\citeyear{junetal081,junetal08});
\citet{smithetal09};
\citet{sainfurrer10};
\citet{christensensain10};
\citet{sainetal10}].
One
important research problem of interest regarding these climate model
outputs is to investigate the cross-correlations across climate model errors
[\citet{tebaldireto07};
\citeauthor{junetal081} (\citeyear{junetal081,junetal08});
\citet{Knutetal10}]. Climate
models are constantly copied and compared, and successful
approximation schemes are frequently borrowed from other climate
models. Therefore, many of the climate models are dependent to some
degree and thus may be expected to have correlated errors. For similar
reasons, we may expect correlations to be higher between the models
developed by the same organization. Indeed,
\citeauthor{junetal081} (\citeyear{junetal081,junetal08}) quantified cross-correlations between
pairs of climate model errors at each spatial location and the results
show that many climate model errors have high correlations and some of
the models developed by the same organizations have even higher
correlated errors. Note that throughout the paper, we use the
terminology ``error'' rather than ``bias'' to describe the discrepancy
between the climate model output and the true climate. The reason is
that we consider the ``error'' as a stochastic process rather than a
deterministic quantity.

In this paper we build a joint statistical model for multiple climate
model errors
that accounts for the spatial dependence of individual models as well
as cross-covariance across different climate models. Our model offers a
nonseparable cross-covariance structure. We work with a climate
variable, surface temperature, from multiple global AOGCM
outputs. We include several covariates such as latitude, land/ocean
effect, and altitude in the mean structure. The marginal and
cross-covariance structure of climate model errors are modeled
using a spatially varying linear model of co-regionalization (LMC). The
resulting covariance structure is
nonstationary and is able to characterize the spatially varying
cross-correlations between multiple climate model errors.
Our modeling approach complements in many ways the previous work
of \citet{junetal08}. \citet{junetal08} used kernel smoothing of the
products of climate model errors to obtain cross-correlations and did
not formally build joint statistical models for multiple
climate model outputs.
One direct product of our model is a continuous surface
map for the cross-covariances. Our Bayesian hierarchical modeling
approach also provides uncertainty measures of the
estimations of the spatially varying cross-covariances, while
it is a challenging task to achieve for the kernel smoothing approach.
In \citet{junetal08} they fit each climate model error separately to a
univariate regression model before obtaining the kernel estimate of
cross-correlations. They only considered land/ocean effect in the
covariance structure.
In our approach, we not only include the land/ocean effect but also
altitude and latitude in the cross-covariance structure of the climate
model errors.
As a~statistical methodology, joint modeling is more efficient
in estimating model parameters. Moreover, our approach is able to
produce spatial prediction or interpolation and thus is potentially
useful as a statistical downscaling technique for multivariate climate
model outputs.

The naive implementation of our modeling approach faces a big
computational challenge because it requires repeated calculations of
quadratic forms in large inverse covariance matrices and determinants
of large covariance matrices,
needed in the posterior distribution and likelihood evaluation.
This is a well-known challenge when dealing with large spatial data
sets. Various methods have been proposed in the literature to overcome
this challenge, including likelihood approximation
[\citet{vecchia1988eam};
\citet{stein2004all};
\citet{fuentes2007all};
\citet{caragea2007asymptotic}],
covariance tapering [\citet{furrer2006cti};
\citet{kaufman2008ctl}],
Gaussian Markov random-field approximation
[\citet{rue2002fgm};
\citet{rue2005gmr}] and
reduced rank approximation [\citet{higdon2002sas};
\citet{wikle1999dra};
\citet{verhoef2004fsm};
\citet{kammann2003gm};
\citet{cressie2008frk};
\citet{banerjee2008gpp}].
Most of these methods focus on univariate processes. One of the
exceptions is the
predictive process approach [\citet{banerjee2008gpp}]. Although the
predictive process approach
is applicable to our multivariate model, it is not a perfect choice
because it usually fails to accurately capture local, small-scale
dependence structures [\citet{Finley2008}].

We develop a new covariance approximation for multivariate processes
that improves the predictive process approach.
Our covariance approximation, called the full-scale approximation,
consists of two parts: The first part is the same as the multivariate
predictive process, which is effective
in capturing large-scale spatial dependence; and the second part
is a sparse covariance matrix that can approximate well the
small-scale spatial dependence, that is, unexplained by the first part.
The complementary form of our covariance approximation enables
the use of reduced-rank and sparse matrix operations and hence
greatly facilitates computation in the application of the Gaussian
process models for large data sets. Although our method is developed to
study multiple climate model errors, it is generally applicable to a
wide range of multivariate Gaussian process models
involving large spatial data sets. The full-scale
approximation has been previously studied for univariate processes in
\citet{sanghuang10}, where covariance tapering was used to generate
the second part of the full-scale approximation. In addition to
the covariance tapering, this paper considers using block diagonal
covariance for the same purpose. Simulation results of model fitting
and prediction using the full-scale approximation show substantial
improvement over the predictive process and the independent blocks
analysis. We also find that using the block diagonal
covariance achieves even faster computation than using covariance
tapering with comparable performance.

Our work contributes to the literature of statistical modeling of
climate model outputs.
Despite high correlations between the errors of some climate models,
it has been commonly assumed that climate model outputs and/or their
errors are independent of each other
[\citet{giorgimearns02};
\citet{tebaldietal05};
\citet{greeneetal06};
\citet{furreretal07};
\citet{smithetal09};
\citet{furrersain09};
\citet{tebaldisanso09}].
Only recently, several authors have attempted to build joint models
for multiple climate model outputs that account for the dependence
across different climate models. For example, \citet{sainfurrer10} used
a multivariate Markov random-field (MRF) approach to analyze 20-year
average precipitation data from five different regional climate model
(RCM) outputs. They presented a method to combine the five RCM outputs
as a weighted average and the weights are determined by the
cross-covariance structure of different RCM outputs. \citet{sainetal10}
built multivariate MRF models for the analysis of two climate
variables, precipitation and temperature, from six ensemble members of
an RCM (three ensemble members for each climate variable). Their model
is designed for what they call \textit{simple ensembles} that resulted
from the perturbed initial conditions of one RCM, not for multiple
RCMs. Unlike our approach, the MRF models are best suited for
lattice data, and it might be difficult for their approach to
incorporate covariates information into the cross-covariance structure.
\citet{christensensain10} used the factor model approach to integrate
multiple RCM outputs. Their analysis is done at each spatial location
independently and, thus, the spatial dependence across grid pixels
is not considered in the analysis. We believe our modeling and
computational method provides a flexible alternative to these
existing approaches in joint modeling of climate model outputs.

The remainder of the paper is organized as follows. In
Section~\ref{secdatamodel} we detail the motivating data set of
climate model errors and present a multivariate spatial process model
to analyze the data. Section \ref{sec3} introduces a covariance approximation
method to address the model fitting issues with large multivariate
spatial data sets. In Section \ref{Sim} we present a simulation study
to explore properties of the proposed computational method. The full analysis
of the climate errors data is then offered in Section \ref{secappresult}. Finally, we
conclude the paper in Section \ref{sec6}.

\section{Data and the multivariate spatial model}
\label{secdatamodel}

\subsection{Data}
\label{secdata}

Our goal is to build a joint statistical model for multiple climate model
errors. By climate model error, we mean the difference between
the\vadjust{\goodbreak}
climate model output and the corresponding observations. We use
surface temperature (unit: K) with global coverage obtained from
the AOGCM simulations as well as the observations.
The observations are provided by the Climate Research Unit (CRU), East
Anglia, and the Hadley Centre, UK MetOffice [\citet{hadcru1};
\citet{hadcru2}].
The data (both climate model outputs and observations) are monthly
averages given on a regular spatial grid (the resolution
is $5^{\circ}\times5^{\circ}$). A list of the climate models used
in this study is given in Table~\ref{list}. These climate models are
developed by different modeling groups worldwide, under the
coordination of the IPCC, and they use common initial conditions. Each
model has their own grid resolution. We consider the 30 year interval
1970--1999 and due to the lack of
observations near the polar area, we only consider temperature data
taken between latitude $45^{\circ}$ S and $72^{\circ}$ N, with the
full longitude ranging from $180^{\circ}$~W to $180^{\circ}$ E.
We still have missing observations at a few pixels
and we impute by taking the average of the spatially neighboring cells
(all eight neighboring cells if all are available).
We study climatological mean state in the sense that we first get a
seasonal average temperature (e.g., an average of the monthly
temperature from
December to February) and then average it over 30 years. We denote this
30 year average of Boreal winter temperature by DJF.

%
\begin{table}
\tabcolsep=3pt
\caption{The names of modeling groups, country, IPCC I.D. and
resolutions of the IPCC model outputs used in the study. The resolution
of the observations is $5^{\circ} \times5^{\circ}$}\label{list}
\begin{tabular*}{\textwidth}{@{\extracolsep{\fill}}lcccc@{}}
\hline
&&&&\textbf{Resolution} \\
&\textbf{Group}&\textbf{Country}&\textbf{IPCC I.D.}&(\textbf{longitude}${}\bolds{\times}{}$\textbf{latitude})\\
\hline
1&US Dept. of Commerce/NOAA/&USA&GFDL-CM2.0&$2.5^{\circ}\times2^{\circ}$\phantom{.}\\
&Geophysical Fluid&&&\\
&Dynamics Laboratory&&&\\
2&US Dept. of Commerce/NOAA/&USA&GFDL-CM2.1&$2.5^{\circ}\times2^{\circ}$\phantom{.}\\
&Geophysical Fluid&&&\\
&Dynamics Laboratory&&&\\
3&Hadley Centre for Climate &UK&UKMO-HadCM3&\phantom{0}$3.79^{\circ}\times
2.47^{\circ}$\\
&Prediction and Research/&&&\\
&Met Office&&&\\
4&Hadley Centre for Climate &UK&UKMO-HadGEM1&$1.875^{\circ}\times1.24^{\circ}$\\
&Prediction and Research/&&&\\
&Met Office&&&\\
5&LASG/Institute of &China&FGOALS-g1.0&$2.81^{\circ}\times3^{\circ}$\phantom{0.}\\
&Atmospheric Physics&&&\\
\hline
\end{tabular*}
\end{table}

As mentioned earlier, we calculate model errors by taking the
difference between climate model outputs and actual observations.
However, as shown in Table~\ref{list}, the climate model outputs and
the observations are at different spatial grid resolutions. Since the
observations have the coarsest grid, we use the bilinear interpolation of
the model output to the observational grid. That is, at each grid pixel
(in the observation grid), we use the weighted average of model outputs
at the four nearest pixels in the model output grid.
Figure~\ref{estimatedmean} displays pictures of the climate model
errors for
the five climate models listed in Table~\ref{list}. First note that
the errors of model 5 are significantly larger in the higher latitudes
and in the Himalayan area compared to the errors of the other
models. The climate model error maps exhibit quite similar patterns
among the models developed by the same group, although the errors for
model 1 in the mid-latitude area of the Northern Hemisphere appear to
be larger in magnitude compared to those of model 2. All the models
have larger errors in the Northern Hemisphere and the errors are large
over the high altitude area and high latitude area.

We focus on climate model errors in this paper. Since the climate model
error is defined as the difference between a climate model's output
and the corresponding observations (we use the observation minus the
model output), we are effectively building a joint model for
multiple climate model outputs with the observations as their means
(although we have to be careful about the sign of the fixed mean part).
It would be tantalizing to build an elaborate joint statistical model
of multiple climate model outputs as well as the observations directly
at their original spatial grid resolutions.
This direct modeling approach would require
a statistical representation of the true climate and of the climate
model outputs, both of which are challenging tasks.
It would be hard to model the true climate in a reasonably
complex way and the characteristics of observation errors and
model errors may not be simple.
Therefore, we do not pursue this direction in this paper.

\subsection{Multivariate spatial regression models}
\label{secmodel}
Let $\mathbf{Y}(\mathbf{s})=(Y_1(\mathbf{s}), \ldots,\break Y_R(\mathbf
{s}))^T $ be an $R \times1$ response
vector along with a $p\times R$ matrix of regressors
$\mathbf{X}(\mathbf{s})=(\mathbf{X}_1(\mathbf{s}), \ldots, \mathbf
{X}_R(\mathbf{s}))$ observed at location
$\mathbf{s}\in D$, where $D$ is the region of interest. For our application,
$Y_i(\mathbf{s})$ represents the error of climate model~$i$ at location
$\mathbf{s}$ for $i=1,\ldots,5$, $\mathbf{X}_i(\mathbf{s})$ the
corresponding covariates, and~$D$ represents the surface of the globe.
We consider the multivariate spatial regression model
%
\begin{equation}\label{MultiSpatialRegressionEqn}
\mathbf{Y}(\mathbf{s}) = \mathbf{X}^{T}(\mathbf{s})\bolds{\beta}+
\mathbf{w}(\mathbf{s})+\bolds{\varepsilon}(\mathbf{s}),\qquad
\mathbf{s}\in D \subseteq\mathbb{R}^d,
\end{equation}
where $\bolds{\beta}=(\beta_{1},\ldots,\beta_{p})^T$ is a $p\times
1$ column
vector of regression coefficients, $\mathbf{w}(\mathbf{s})$ is a multivariate
spatial process whose detailed specification is given below, and
the process
$\bolds{\varepsilon}(\mathbf{s})=(\varepsilon_{1}(\mathbf{s}),\ldots
,\varepsilon_{R}(\mathbf{s}))^{T}$
models the measurement error for the responses. The measurement
error process is typically assumed to be spatially independent,
and at each location,
$\bolds{\varepsilon}(\mathbf{s}) \sim\operatorname{MVN}(\mathbf{0},
\bolds{\Sigma}_{\bolds{\varepsilon}})$,
where MVN stands for the multivariate normal distribution,
and $\bolds{\Sigma}_{\bolds{\varepsilon}}$ is an $R\times R$
covariance matrix.

As a crucial part of model \eqref{MultiSpatialRegressionEqn}, the
spatial process $\mathbf{w}(\mathbf{s}) = (w_1(\mathbf{s}),
\ldots,\break
w_R(\mathbf{s}))^T$ captures
dependence both within measurements at a given site and across
the sites. We model $\mathbf{w}(\mathbf{s})$ as an $R$-dimensional zero-mean
multivariate Gaussian process:
$\mathbf{w}(\mathbf{s})\sim\operatorname{MVN}(\mathbf{0}, \Gamma
_{\mathbf{w}}(\cdot,\cdot))$, where
the cross-covariance matrix function of $\mathbf{w}(\mathbf{s})$ is
defined as
$\Gamma_{\mathbf{w}}(\mathbf{s},\mathbf{s}')=[\operatorname
{Cov}(w_{r}(\mathbf{s}),w_{r'}(\mathbf{s}'))]_{r, r'=1}^{R}$.
For any integer $n$ and any collection of sites
$\mathcal{S}=(\mathbf{s}_1, \ldots, \mathbf{s}_n)$, we denote the
multivariate
realizations of $\mathbf{w}(\mathbf{s})$ at $\mathcal{S}$ as an
$nR\times1$ vector
$\mathbf{w}=(\mathbf{w}^{T}(\mathbf{s}_{1}),\ldots,\break\mathbf
{w}^{T}(\mathbf{s}_{n}))^{T}$, which follows an
$nR\times1$ multivariate normal distribution $\mathbf{w}\sim
\operatorname{MVN}(\mathbf{0},\bolds{\Sigma}_{\mathbf{w}})$, where
$\bolds{\Sigma}_{\mathbf{w}}=[\Gamma_{\mathbf{w}}(\mathbf
{s}_{i},\mathbf{s}_{j})]_{i, j=1}^{n}$ is an
$nR\times nR$ matrix that can be partitioned as an $n\times n$
block matrix with the $(i, j)$th block being the $R\times R$
cross-covariance matrix $\Gamma_{\mathbf{w}}(\mathbf{s}_i, \mathbf{s}_j)$.
In the multivariate setting, we require a valid cross-covariance
function such that the resultant $nR\times nR$ covariance matrix,~$\bolds{\Sigma}_{\mathbf{w}}$,
is positive definite.

There have been many works on the construction of flexible
cross-covarian\-ce functions
[\citet{mardia1993multivariate};
\citet{higdon1999nss};
\citet{gaspari1999construction};
\citet{higdon2002sas};
\citet{verhoef2004fsm};
\citet{gelfand2004nmp};
\citet{majumdar2007multivariate};
\citet{Gneiting2009};
\citet{jun09};
\citet{apanagenton}].
The model in \citet{mardia1993multivariate} assumes separable
cross-covarian\-ce functions.
\citet{higdon2002sas} employs discrete approximation to the kernel
convolution based on a set of prespecified square-integrable kernel
functions. The model in \citet{Gneiting2009} is isotropic and the
covariance model requires the same spatial range parameters when there
are more than two processes.
The model of \citet{apanagenton} is developed to model stationary
processes and its extension to nonstationary processes is not
obvious.

We adopt the LMC approach [\citet{wackernagel2003mgi};
\citet{gelfand2004nmp}]
which has recently gained popularity in multivariate spatial
modeling due to its richness in structure and feasibility in
computation. Suppose that $\mathbf{U}(\mathbf{s})=[U_{q}(\mathbf
{s})]_{q=1:Q}$ is a
$Q\times1$ process with
each $U_{q}(\mathbf{s})$ independently modeled as a univariate spatial
process with mean zero, unit variance and correlation function
$\rho_{q}(\cdot,\cdot)$. The cross-covariance function of $\mathbf
{U}(\mathbf{s})$
is a diagonal matrix that can be written as
$\Gamma_{\mathbf{u}} (\mathbf{s},\mathbf{s}')=\bigoplus_{q=1}^{Q}\rho
_{q}(\mathbf{s},\mathbf{s}')$,
where $\bigoplus$ is the direct sum matrix operator
[e.g., \citet{harville2008matrix}].
The LMC approach assumes that $\mathbf{w}(\mathbf{s})=\mathbf
{A}(\mathbf{s})\mathbf{U}(\mathbf{s})$, where
$\mathbf{A}(\mathbf{s})$ is an $R\times Q$ transformation matrix,
that is, nonsingular for all $\mathbf{s}$. For identifiability purposes and
without loss of generality, $\mathbf{A}(\mathbf{s})$ can be taken to
be a
lower-triangular matrix. It follows that the constructed
cross-covariance function for $\mathbf{w}(\mathbf{s})$ under this
model is
$\Gamma_{\mathbf{w}}(\mathbf{s},\mathbf{s}')=\mathbf{A}(\mathbf
{s})\Gamma_{\mathbf{u}}(\mathbf{s},\mathbf{s}')\mathbf
{A}^T(\mathbf{s}')$.
An alternative expression for the cross-covariance function is
$\Gamma_{\mathbf{w}}(\mathbf{s}, \mathbf{s}') =
\sum_{q=1}^{Q}\mathbf{a}_{q}(\mathbf{s})\mathbf{a}_{q}^{T}(\mathbf
{s}')\rho_{q}(\mathbf{s},\mathbf{s}')$,
where $\mathbf{a}_{q}(\mathbf{s})$ is the $q$th column vector of
$\mathbf{A}(\mathbf{s})$.
The cross-covariance matrix for the realizations $\mathbf{w}$ at
$n$ locations $\mathcal{S}$ can be written as a block partitioned matrix~$\bolds{\Sigma}_{\mathbf{w}}$ with $n\times n$ blocks, whose
$(i,j)$th block is
$\mathbf{A}(\mathbf{s}_{i})\Gamma_{\mathbf{u}}(\mathbf{s}_i,
\mathbf{s}_j)\mathbf{A}^{T}(\mathbf{s}_{j})$.
We can express $\bolds{\Sigma}_{\mathbf{w}}$ as
%
\begin{equation}\label{eqcovw}
\bolds{\Sigma}_{\mathbf{w}} = \Biggl[\bigoplus_{i=1}^{n}\mathbf{A}(\mathbf
{s}_i)\Biggr]\Biggl[\bigoplus_{q=1}^{Q} \rho
_{q}(\mathbf{s}_i,\mathbf{s}_j)\Biggr]_{i, j=1}^{n}\Biggl[\bigoplus
_{i=1}^{n}\mathbf{A}^{T}(\mathbf{s}_i)\Biggr]=\mathcal{A}
\bolds{\Sigma}_{\mathbf{u}}\mathcal{A}^{T},
\end{equation}
where $\mathcal{A}$ is a block-diagonal matrix with $n\times n$ blocks whose
$i$th diagonal block is
$\mathbf{A}(\mathbf{s}_i)$, $\bigoplus_{q=1}^{Q}\rho_{q}(\mathbf
{s}_i, \mathbf{s}_j)$ is a $Q \times Q$
diagonal matrix with $\rho_{q}(\mathbf{s}_i , \mathbf{s}_j)$'s as
its diagonals, and
$\bolds{\Sigma}_{\mathbf{u}}$ is an $(n\times n)$-block partitioned
matrix with
$\Gamma_{\mathbf{u}}(\mathbf{s}_i, \mathbf{s}_j)$ as its $(i,j)$th block.

For our climate model application, we utilize the multivariate model
\eqref{MultiSpatialRegressionEqn}. Here, we have $R=5$ since five
climate model errors are considered. In the LMC model, the number of
latent processes, $Q$, can take a value from $1$ to $R$. When $Q<R$,
the multivariate process is represented in a lower dimensional space
and dimensionality reduction is achieved.
In this paper, our goal is to obtain a rich, constructive
class of multivariate spatial process models and, therefore, we assume
a full rank LMC model with $Q = R =5$. We do not include the spatial
nugget effect~$\bolds{\varepsilon}$ in the model. For the regression
mean,~$\mathbf{X}
^T(\mathbf{s}) \bolds{\beta}$, we use $p=8$ covariates: Legendre
polynomials [\citet{abramowitz1964handbook}] in latitude of order
$0$ to $4$ with sine of latitude as their arguments, an indicator of
land/ocean (we give 1 if the domain is over the land and 0 otherwise),
longitude (unit: degree), and the altitude (altitude
is set to be zero over the ocean). We scale the altitude variable
(unit: m) by
dividing it by 1000 to make its range comparable to other covariates.
Our covariates specification is similar to \citet{sainetal10}
except that we use multiple terms for latitude effects in order to
have enough flexibility to capture the dependence of the process
on the entire globe.

For each location $\mathbf{s}$, we model $\mathbf{A}(\mathbf{s})$ as
a $5 \times5$ lower
triangular matrix. Two model specifications for $\mathbf{A}(\mathbf
{s})$ are
considered. In one specification, we assume the linear
transformation to be independent of space, that is, $\mathbf
{A}(\mathbf{s})=\mathbf{A}$. Thus,
we have $\mathbf{A}(\mathbf{s})=[a_{ij}]_{i,j =1:5}$ with $a_{ij}$
being nonzero
constants for $1\leq i \leq j \leq5$ and $a_{ij}=0$ for $i>j$. To
avoid an identifiability problem, we let $a_{ii}>0$ for $i=1,\ldots,5$.
If the $U_{q}(\mathbf{s})$ are stationary, this specification results
in a
stationary cross-covariance structure. In particular, if the
$U_{q}(\mathbf{s}
)$ are identically distributed, this specification results in a separable
cross-covariance structure similarly to \citet{sainfurrer10}.
In the second specification, we assume that $\mathbf{A}(\mathbf{s})$
vary over
space. Specifically, the $(i,j)$th entry of $\mathbf{A}(\mathbf{s})$,
$a_{ij}(\mathbf{s})$, is modeled as $a_{ij}(\mathbf{s})=\mathbf
{X}_{A}^{T}(\mathbf{s})\bolds{\eta}_{ij}$
for $1\leq i\leq j\leq5$, where $\mathbf{X}_{A}(\mathbf{s})$ is a~$5
\times1$
covariate vector at location $\mathbf{s}$ and consists of Legendre polynomials
in latitude of order
$0$ to $2$, an indicator of land/ocean, and the scaled altitude.
The dimension of $\bolds{\eta}_{ij}$ is $5 \times1$.
This specification induces nonstationary and nonseparable
cross-covariance structure. Note that, similar to the first
specification, to avoid an identifiability problem, we let
$a_{ii}(\mathbf{s})=|\mathbf{X}_{A}^T(\mathbf{s})\bolds{\eta
}_{ii}|$ for $i=1,\ldots,5$.
\citet{gelfand2004nmp} proposed to model each element of~$\mathbf
{A}(\mathbf{s})$ as
a spatial process, but in practice such an approach
is usually computationally too demanding to be used for large scale problems.
Moreover, there might be identifiability problems if we do not
constrain some of the elements in $\mathbf{A}(\mathbf{s})$. Thus, we
do not
consider this option in our paper.
For the correlation function of each latent process $U_{q}(\mathbf
{s})$, we
consider a Mat\'{e}rn correlation function on a
sphere, where the chordal distance is used. For any two
locations on the globe $(L_i,l_i), i=1,2$, where $L_i$ and $l_i$
denote latitude and longitude, respectively, the chordal distance
is defined as\looseness=-1
\begin{eqnarray*}
&&\operatorname{ch}((L_1,l_1),(L_2, l_2))\\
&&\qquad=2R_{\mathrm{earth}}\biggl\{\sin
^2\biggl(\frac
{L_1-L_2}{2}\biggr)+\cos L_1 \cos L_2\sin^2\biggl(\frac{l_1-l_2}{2}\biggr)\biggr\}^{1/2}.
\end{eqnarray*}\looseness=0
Here $R_{\mathrm{earth}}$ is the radius of the Earth. \citet{junetal08}
showed that
the maximum likelihood estimates of the smoothness parameters for all
of the climate models in Table~\ref{list} (except for model 4) are
close to 0.5. Therefore, we fix the smoothness parameter values for all
the processes in $\mathbf{U}$ to be 0.5 and this is the same as using an
exponential correlation function for each $U_{q}(\mathbf{s})$.

\section{Model implementation and covariance approximation}\label{sec3}

\subsection{Model fitting}
\label{ModelFit}

We adopt a Bayesian approach that specifies prior distributions on the
parameters. Posterior inference for the model parameters is
implemented by model fitting with Gibbs samplers
[\citet{gelfand1990sba}] and Metropolis--Hastings updating
[\citet{gelman2004bda}]. We set~$\bolds{\beta}$ to have a \mbox{$p$-dimensional}
multivariate normal distribution prior with large variances.
The prior assignment for the covariance parameters for each
$\mathbf{U}_i$ depends upon the specific choice
of the correlation functions. In general, the spatial range parameters
are weakly identifiable, and, hence, reasonably informative priors are
needed for satisfactory MCMC behavior. We set prior distributions
for the spatial range parameters relative to the size of
their domains, for instance, by setting the
prior means to reflect one's prior belief about the
practical spatial range of the data.
For the LMC setting with a constant $\mathbf{A}$, we may
assign truncated normal priors with positive value support or inverse
gamma priors with infinite variances for the diagonal entries, and
normal priors for other entries. For the spatially varying LMC
setting, the priors for the coefficients $\bolds{\eta}_{ij}$ are normal
distributions with large variances.

Given $n$ locations in the set $\mathcal{S} = \{\mathbf{s}_1,\ldots
,\mathbf{s}_n\}$,
the realization of the response vector at these locations can be
collectively denoted as $\mathbb{Y}= (\mathbf{Y}(\mathbf{s}_1)^T,\break
\ldots,
\mathbf{Y}(\mathbf{s}_n)^T)^T$, and the corresponding matrix of
covariates is
$\mathbb{X}= (\mathbf{X}(\mathbf{s}_1), \ldots,\break \mathbf{X}(\mathbf
{s}_n))^T$.
The data likelihood can be obtained easily from the fact
that $\mathbb{Y}\sim\operatorname{MVN}(\mathbb{X}\bolds{\beta
},\bolds{\Sigma}_{\mathbf{w}}
+\mathbf{I}_{n}\otimes\bolds{\Sigma}_{\bolds{\varepsilon}})$, where
$\bolds{\Sigma}_{\mathbf{w}}$ is given
by \eqref{eqcovw}.
Generically denoting the set of all model parameters by $\Omega$,
the MCMC method is used to draw samples of the model parameters
from the posterior: $p(\Omega|\mathbb{Y}) \propto
P(\Omega)P(\mathbb{Y}|\Omega)$. Assuming the prior distribution of~$\bolds{\beta}$
is $\operatorname{MVN}(\bolds{\mu}_{\bolds{\beta}},\bolds{\Sigma
}_{\bolds{\beta}})$, the posterior samples
of $\bolds{\beta}$ are
updated from its full conditional
$\operatorname{MVN}(\bolds{\mu}_{\bolds{\beta}|\cdot},\bolds
{\Sigma}_{\bolds{\beta}|\cdot})$, where
$\bolds{\Sigma}_{\bolds{\beta}|\cdot}=[\bolds{\Sigma}_{\bolds
{\beta}}^{-1}
+\mathbb{X}^{T}(\bolds{\Sigma}_{\mathbf{w}}+\mathbf{I}_{n}\otimes
\bolds{\Sigma}_{\bolds{\varepsilon}})^{-1}\mathbb{X}]^{-1}$, and
$\bolds{\mu}_{\bolds{\beta}|\cdot} =
\bolds{\Sigma}_{\bolds{\beta}|\cdot}\mathbb{X}^{T}(\bolds{\Sigma
}_{\mathbf{w}} +\mathbf{I}_{n}\otimes
\bolds{\Sigma}_{\bolds{\varepsilon}})^{-1}\mathbb{Y}$.
All the remaining parameters have to be updated using
Metropolis--Hastings steps.

Spatial interpolation in the multivariate case allows one to better
estimate one variable at an unobserved location by borrowing
information from co-located variables. It is also called ``cokriging''
in geostatistics. The multivariate spatial regression model provides
a natural way to do cokriging. For example, under the Bayesian inference
framework, the predictive distribution for
$\mathbf{Y}(\mathbf{s}_0)=[Y_{i}(\mathbf{s}_{0})]_{i=1:p}$ at a new
location $\mathbf{s}_{0}$ is a
Gaussian distribution with
\[
\mathrm{E}[\mathbf{Y}(\mathbf{s}_0)|\Omega,\mathbb{Y}] = \mathbf
{X}^{T}(\mathbf{s}_0)\bolds{\beta}+
\mathbf{h}(\mathbf{s}_0)(\bolds{\Sigma}_{\mathbf{w}} + \mathbf
{I}_{n}\otimes\bolds{\Sigma}_{\bolds{\varepsilon}
})^{-1}(\mathbb{Y}-
\mathbb{X}\bolds{\beta})
\]
and
\[
\operatorname{Cov}[\mathbf{Y}(\mathbf{s}_0)|\Omega,\mathbb{Y}] =
\Gamma_{\mathbf{w}}(\mathbf{s}_0,\mathbf{s}_{0}) +
\bolds{\Sigma}_{\bolds{\varepsilon}} -
\mathbf{h}(\mathbf{s}_0)(\bolds{\Sigma}_{\mathbf{w}} + \mathbf
{I}_{n}\otimes\bolds{\Sigma}_{\bolds{\varepsilon}})^{-1}\mathbf{h}
^{T}(\mathbf{s}_0),
\]
where $\mathbf{h}(\mathbf{s}_0)=[\Gamma_{\mathbf{w}}(\mathbf
{s}_0,\mathbf{s}_{i})]_{i=1:n}$ is the
$R\times nR$ cross-covariance matrix between $\mathbf{w}(\mathbf
{s}_0)$ and
$\{\mathbf{w}(\mathbf{s}_i), i=1, \ldots,n\}$.

The computationally demanding part in the model fitting is to
calculate the quadratic form of the inverse of the $nR \times nR$ matrix
$\bolds{\Sigma}_{\mathbf{w}}+\mathbf{I}_{n}\otimes\bolds{\Sigma
}_{\bolds{\varepsilon}}$, whose computational
complexity is of the order $O((nR)^3)$. This computational issue
is often referred
to as ``the big n problem.'' In our climate model application, we
have 1,656 locations for each of the five climate models, that is,
$n=1\mbox{,}656$ and $R=5$. Although the inversion of the matrix can be
facilitated by the Cholesky decomposition and linear solvers, computation
remains expensive when $nR$ is big, especially for the spatially
varying LMC models which involve a relatively large number of
parameters and hence require multiple likelihood evaluations at each
MCMC iteration. We introduce covariance approximation below as a way
to gain computational speedup for the implementation of the LMC
models.

\subsection{Predictive process approximation}
\label{secpred-process}

In this subsection we review the multivariate predictive process
approach by \citet{banerjee2008gpp} and point out its drawbacks to
motivate our new covariance approximation method.
The predictive process models consider a fixed set of
``knots'' $\mathcal{S}^{\ast} =
\{\mathbf{s}^{\ast}_1,\ldots,\break\mathbf{s}^{\ast}_{m}\}$ that are
chosen from the study
region. The Gaussian process $\mathbf{w}(\mathbf{s})$ in model~(\ref
{MultiSpatialRegressionEqn}) yields an $mR\time1$ random vector
$\mathbf{w}
^{\ast}=[\mathbf{w}(\mathbf{s}_{i}^{\ast})]_{i=1}^{m}$ over $\mathcal{S}^{\ast}$. The
Best Linear Unbiased Predictor (BLUP) of $\mathbf{w}(\mathbf{s})$ at
any fixed site
$\mathbf{s}$ based on~$\mathbf{w}^{\ast}$ is
given by $\tilde{\mathbf{w}}(\mathbf{s}) = E\{\mathbf{w}(\mathbf
{s})|\mathbf{w}^{\ast}\}$. Being a
conditional expectation, it immediately follows that $\tilde{\mathbf
{w}}(\mathbf{s})$
is an optimal predictor of $\mathbf{w}(\mathbf{s})$ in the sense that
it minimizes
the mean squared prediction error $E\{\|\mathbf{w}(\mathbf{s})-
\mathbf{f}(\mathbf{w}^*)\|^2\}$
over all square-integrable (vector-valued) functions $\mathbf
{f}(\mathbf{w}^*)$
for Gaussian processes, and over all linear functions without the
Gaussian assumption. \citet{banerjee2008gpp} refer to $\tilde{\mathbf
{w}}(\mathbf{s})$
as the predictive process derived from the parent process $\mathbf
{w}(\mathbf{s})$.

Since the parent process $\mathbf{w}(\mathbf{s})$ is an
$R$-dimensional zero-mean
multivariate Gaussian process with the cross-covariance function
$\Gamma_{\mathbf{w}}(\mathbf{s},\mathbf{s}') = \operatorname
{Cov}(\mathbf{w}(\mathbf{s}),\break\mathbf{w}(\mathbf{s}'))$,
the multivariate predictive process has a
closed-form expression
%
\begin{equation}\label{MultivPredProcessEqn}
\tilde{\mathbf{w}}(\mathbf{s}) =
\operatorname{Cov}(\mathbf{w}(\mathbf{s}),\mathbf{w}^{\ast
})\operatorname{Var}^{-1}(\mathbf{w}^{\ast})\mathbf{w}^{\ast} =
\mathcal{C}_{\mathbf{w}}(\mathbf{s},\mathcal{S}^{\ast};\bolds
{\theta})\mathcal{C}_{\mathbf{w}}^{\ast-1}(\bolds{\theta})\mathbf{w}
^{\ast},
\end{equation}
where $\mathcal{C}_{\mathbf{w}}(\mathbf{s},\mathcal{S}^{\ast
};\bolds{\theta})
=[\Gamma_{\mathbf{w}}(\mathbf{s},\mathbf{s}^{\ast}_1;\bolds
{\theta}),\ldots,\Gamma_{\mathbf{w}}
(\mathbf{s},\mathbf{s}^{\ast}_m;\bolds{\theta})]$
is an $R\times mR$ cross-\break covariance matrix between $\mathbf{w}(\mathbf{s})$
and $\{\mathbf{w}(\mathbf{s}^*), \mathbf{s}^*\in\mathcal{S}^{\ast
}\}$, and
$\mathcal{C}_{\mathbf{w}}^{\ast}(\bolds{\theta})
=[\Gamma_{\mathbf{w}}(\mathbf{s}^{\ast}_i,\mathbf{s}^{\ast
}_j;\break \bolds{\theta})]_{i,j=1}^{m}$
is the $mR\times mR$ cross-covariance matrix of
$\mathbf{w}^{\ast}=[\mathbf{w}(\mathbf{s}^{\ast}_i)]_{i=1}^{m}$
[see, e.g., \citet{banerjee2004hma}].

The multivariate predictive process $\tilde{\mathbf{w}}(\mathbf{s})$
is still a zero
mean Gaussian process, but now with a fixed rank cross-covariance
function given by
$\Gamma_{\tilde{\mathbf{w}}}(\mathbf{s},\mathbf{s}') =
\mathcal{C}_{\mathbf{w}}(\mathbf{s},\mathcal{S}^{\ast};\bolds
{\theta})\mathcal{C}_{\mathbf{w}}^{\ast
-1}(\bolds{\theta})\mathcal{C}_{\mathbf{w}}^{T}(\mathbf
{s}',\mathcal{S}^{\ast};\bolds{\theta})$.
Let $\tilde{\mathbf{w}}=[\tilde{w}(\mathbf{s}_i)]_{i=1}^{n}$ be the
realization of
$\tilde{\mathbf{w}}(\mathbf{s})$ at the set $\mathcal{S}$ of the
observed locations.
It follows that $\tilde{\mathbf{w}}\sim\operatorname{MVN}(\mathbf
{0},\bolds{\Sigma}_{\tilde{\mathbf{w}}})$,
where $\bolds{\Sigma}_{\tilde{\mathbf{w}}}=\mathcal{C}_{\mathbf
{w}}(\mathcal{S},\mathcal{S}^{\ast};\bolds{\theta})\mathcal{C}
_{\mathbf{w}}^{\ast-1}(\bolds{\theta})\mathcal{C}_{\mathbf
{w}}^{T}(\mathcal{S},\mathcal{S}^{\ast};\bolds{\theta})$, and
$\mathcal{C}_{\mathbf{w}}(\mathcal{S},\mathcal{S}^{\ast};\bolds
{\theta})$ is an $nR\times mR$
matrix that can be partitioned as an $n\times1$ block matrix whose
$i$th block is given by $\mathcal{C}_{\mathbf{w}}(\mathbf
{s}_i,\mathcal{S}^{\ast};\bolds{\theta})$.

Replacing $\mathbf{w}(\mathbf{s})$ in \eqref
{MultiSpatialRegressionEqn} with the
fixed rank approximation $\tilde{\mathbf{w}}(\mathbf{s})$, we obtain
the following
multivariate regression model:
%
\begin{equation}\label{MultiSpatialRegressionEqn-pred}
\mathbf{Y}(\mathbf{s}) = \mathbf{X}^{T}(\mathbf{s})\bolds{\beta}+
\tilde{\mathbf{w}}(\mathbf{s}) + \bolds{\varepsilon}(\mathbf{s}),
\end{equation}
which is called the predictive process approximation of
model \eqref{MultiSpatialRegressionEqn}. Based on this approximation,
the data likelihood can be obtained using
$\mathbb{Y}\sim\operatorname{MVN}(\mathbb{X}\bolds{\beta
},\break\mathcal{C}_{\mathbf{w}}(\mathcal{S}, \mathcal{S}^{\ast
};\bolds{\theta})\mathcal{C}_{\mathbf{w}}^{\ast-1}(\bolds{\theta
})\mathcal{C}_{\mathbf{w}}^{T}(\mathcal{S},\mathcal{S}^{\ast
};\bolds{\theta}))+\mathbf{I}_{n}\otimes\bolds{\Sigma}_{\bolds
{\varepsilon}})$.
When the number of knots $m$ is chosen to be substantially
smaller than $n$, computational gains are achieved since
the likelihood evaluation that initially involves inversion of
$nR\times nR$ matrices can be done by inverting much smaller
$mR\times mR$ matrices.

The multivariate predictive process is an attractive reduced rank approach
to deal with large spatial data sets. It encompasses a very flexible
class of spatial cross-covariance models since any given multivariate
spatial Gaussian process with a valid cross-covariance function would
induce a multivariate predictive process.
Since the predictive process is still a valid Gaussian process, the
inference and prediction schemes for multivariate Gaussian spatial
process models that we described in Section \ref{ModelFit}
can be easily implemented here.

However, the predictive process models share one common problem with
many other reduced rank approaches: They generally fail to capture
local/small scale dependence accurately
[\citet{stein2008mal};
\citet{Finley2008};
\citet{banerjee2010hierarchical}] and thus
lead to biased parameter estimations and errors in prediction.

To see the problems with the multivariate spatial process $w(\mathbf
{s})$ in
\eqref{MultiSpatialRegressionEqn},
we consider the following decomposition of the parent process:
%
\begin{equation}\label{eqadd-model}
\mathbf{w}(\mathbf{s}) = \tilde{\mathbf{w}}(\mathbf{s}) + \bigl(\mathbf
{w}(\mathbf{s})-\tilde{\mathbf{w}}(\mathbf{s})\bigr).
\end{equation}
We call $\mathbf{w}(\mathbf{s}) - \tilde{\mathbf{w}}(\mathbf{s})$
the \textsl{residual process}.
The decomposition in \eqref{eqadd-model} immediately implies a
decomposition of the covariance function of the process~$\mathbf
{w}(\mathbf{s})$:
%
\begin{equation}\label{eqadd-covmodel}
\Gamma_{\mathbf{w}}(\mathbf{s},\mathbf{s}') = \Gamma_{\tilde
{\mathbf{w}}}(\mathbf{s},\mathbf{s}')+\bigl(\Gamma_{\mathbf{w}}(\mathbf{s}
,\mathbf{s}')-\Gamma_{\tilde{\mathbf{w}}}(\mathbf{s},\mathbf{s}')\bigr),
\end{equation}
where $\Gamma_{\mathbf{w}}(\mathbf{s},\mathbf{s}')-\Gamma_{\tilde
{\mathbf{w}}}(\mathbf{s},\mathbf{s}')$ is
the cross-covariance function of the residual process
$\mathbf{w}(\mathbf{s}) - \tilde{\mathbf{w}}(\mathbf{s})$.
Note that for any arbitrary set of $n$ locations $\mathcal{S}$,
$\bolds{\Sigma}_{\mathbf{w}}-\bolds{\Sigma}_{\tilde{\mathbf{w}}} =
[\Gamma_{\mathbf{w}}(\mathbf{s}_{i},\mathbf{s}_{j})]_{i, j=1}^{n}
- [\Gamma_{\tilde{\mathbf{w}}}(\mathbf{s}_{i},\mathbf{s}_{j})]_{i,
j=1}^{n}$
is the conditional
variance--covariance matrix of $\mathbf{w}$ given $\mathbf{w}^{\ast
}$, and hence
a nonnegative definite matrix.
Using the multivariate predictive process to approximate $\mathbf
{w}(\mathbf{s})$, we
discard the residual process $\bolds{\Sigma}_{\mathbf{w}}-\bolds
{\Sigma}_{\tilde{\mathbf{w}}}$
entirely and thus fail to capture the dependence it carries. This is
indeed the fundamental issue that leads to biased estimation in the
model parameters.

To understand and illustrate the issue with the
predictive process due to ignoring the residual process, we consider a
univariate stationary Gaussian process and we remark that the
multivariate predictive process shares the same problems. Assume the
covariance function of the parent process is $\mathbf{C}_{w}(\mathbf
{s}, \mathbf{s}')
=\sigma^2\rho_{w}(\mathbf{s}, \mathbf{s}')$, where $\rho
_{w}(\mathbf{s}, \mathbf{s}')$ is the
correlation function and $\sigma^2$ is the variance, that is, constant
over space. Assume $\tau^2$ is the nugget variance. The variance of
the corresponding predictive process at location $\mathbf{s}$ is
given by $\sigma_{\tilde{w}}^{2}(\mathbf{s})=\sigma^2\rho
_{w}^{T}(\mathbf{s},\mathcal{S}^{\ast
})\rho_{w}^{-1}(\mathcal{S}^{\ast},\mathcal{S}^{\ast})\rho
_{w}(\mathbf{s},\mathcal{S}^{\ast})$, where
$\rho_{w}(\mathbf{s},\mathcal{S}^{\ast})$ is the correlation vector
between $w(\mathbf{s})$
and $\{w(\mathbf{s}^*), \mathbf{s}^* \in\mathcal{S}^{\ast}\}$, and
$\rho_{w}(\mathcal{S}^{\ast},\mathcal{S}^{\ast})$ is the correlation
matrix of the realizations of $w(\mathbf{s})$ at the knots in the set
$\mathcal{S}
^{\ast}$.
From the nonnegative definiteness of the residual covariance function, we
obtain the inequality $\sigma^2_{w}(\mathbf{s}) \geq
\sigma_{\tilde{w}}^2(\mathbf{s})$. Equality holds when $\mathbf{s}$
belongs to
$\mathcal{S}^*$. Therefore, the predictive process
produces a lower spatial variability.
\citet{banerjee2010hierarchical} proved that there are systematic
upward biases in likelihood-based estimates of the spatial variance
parameter $\sigma^2$ and the nugget variance $\tau^2$ using the
predictive process model as compared to the parent model. Indeed, the
simulation results in \citet{Finley2008} and \citet{banerjee2010hierarchical}
showed that both $\sigma^2$ and~$\tau^2$
are significantly overestimated especially when the number of knots is small.
Our simulation study to be presented in Section~\ref{Sim} also shows
that predictive processes produce biased estimations of model parameters
in the multivariate spatial case.

\subsection{Full-scale covariance approximation}
\label{secapprox}
Motivated by the fact that discarding the residual process
$\tilde{\bolds{\varepsilon}}(\mathbf{s}) = \mathbf{w}(\mathbf
{s})-\tilde{\mathbf{w}}(\mathbf{s})$
is the main cause of the problem associated with the multivariate predictive
process, we seek to complement the multivariate predictive process by
adding a component that approximates the residual cross-covariance
function while still maintaining computational efficiency.
We approximate the residual cross-covariance function by
%
\begin{equation}\label{eqtapering}
\Gamma_{\tilde{\bolds{\varepsilon}}}(\mathbf{s},\mathbf{s}')=[\Gamma
_{\mathbf{w}}(\mathbf{s},\mathbf{s}';\bolds{\theta})-
\Gamma_{\tilde{\mathbf{w}}}(\mathbf{s},\mathbf{s}';\bolds{\theta
})]\circ\mathcal{K}(\mathbf{s},\mathbf{s}'),
\end{equation}
where $\circ$ denotes the Schur product (or entrywise product) of matrices,
and the matrix-valued function $\mathcal{K}(\mathbf{s},\mathbf
{s}')$, referred to as the
modulating function,
has the property of being a zero matrix for a large proportion of possible
spatial location pairs $(\mathbf{s}, \mathbf{s}')$. The zeroing
property of
the modulating function implies that the resulting cross-covariance
matrix is sparse and, thus, sparse matrix algorithms are readily
applicable for fast computation. We will introduce below
modulating functions that have zero value when $\mathbf{s}$ and
$\mathbf{s}'$
are spatially farther apart. For such choices of the modulating
function, the effect of multiplying a modulating function
in \eqref{eqtapering} is expected to be small, since the residual
process mainly captures the small scale spatial variability.

Combining \eqref{eqadd-covmodel} with \eqref{eqtapering},
we obtain the following approximation of the cross-covariance function:
%
\begin{equation}\label{eqfull-scale}
\Gamma_{\mathbf{w}}^\dag(\mathbf{s}, \mathbf{s}') = \Gamma
_{\tilde{\mathbf{w}}}(\mathbf{s}, \mathbf{s}')
+ \Gamma_{\tilde{\bolds{\varepsilon}}}(\mathbf{s},\mathbf{s}').
\end{equation}
Note that the first part of the cross-covariance approximation,
$\Gamma_{\tilde{\mathbf{w}}}$, is the result of the predictive process
approximation and should capture well the large scale spatial
dependence, while the second part, $\Gamma_{\tilde{\bolds{\varepsilon}}}$,
should capture well the small scale spatial dependence.
We refer to \eqref{eqfull-scale} as the \textsl{full-scale
approximation} (FSA) of the original cross-covariance function.

Using the FSA, the covariance matrix of the data $\mathbb{Y}$ from
model \eqref{MultiSpatialRegressionEqn} is approximated by
%
\begin{equation}\label{eqdatacovapp}
\bolds{\Sigma}_{\mathbb{Y}} =
\bolds{\Sigma}_{\tilde{\mathbf{w}}}+\bolds{\Sigma}_{\tilde
{\bolds{\varepsilon}}}+\mathbf{I}_{n}\otimes\bolds{\Sigma}
_{\bolds{\varepsilon}},
\end{equation}
where $\bolds{\Sigma}_{\tilde{\mathbf{w}}}=\mathcal{C}_{\mathbf
{w}}(\mathcal{S},\mathcal{S}^{\ast};\bolds{\theta})\mathcal{C}
_{\mathbf{w}}^{\ast
-1}(\bolds{\theta})\mathcal{C}_{\mathbf{w}}^{T}(\mathcal
{S},\mathcal{S}^{\ast};\bolds{\theta})$,
and $\bolds{\Sigma}_{\tilde{\bolds{\varepsilon}}}=[\Gamma_{\tilde
{\bolds{\varepsilon}}}(\mathbf{s}_{i},\mathbf{s}_{j};\bolds{\theta}
)]_{i,j=1}^{n}$.
The structure of the covariance matrix \eqref{eqdatacovapp}
allows efficient computation of the quadratic form of its inverse
and its determinant.
Using the Sherman--Woodbury--Morrison formula, we see that
the inverse of $\bolds{\Sigma}_{\mathbb{Y}}$ can be computed by
%
\begin{eqnarray}\label{eqswm-inv}
(\bolds{\Sigma}_{\tilde{\mathbf{w}}}\,{+}\,\bolds{\Sigma}_{\tilde
{\bolds{\varepsilon}}}\,{+}\,\mathbf{I}_{n}\otimes\bolds{\Sigma}
_{\bolds{\varepsilon}})^{-1}
&\,{=}\,&(\bolds{\Sigma}_{\tilde{\bolds{\varepsilon}}}\,{+}\,\mathbf
{I}_{n}\,{\otimes}\,\bolds{\Sigma}_{\bolds{\varepsilon}})^{-1}\,{-}\,
(\bolds{\Sigma}_{\tilde{\bolds{\varepsilon}}}\,{+}\,
\mathbf{I}_{n}\,{\otimes}\,
\bolds{\Sigma}_{\bolds{\varepsilon}})^{-1}
\mathcal{C}_{\mathbf{w}}(\mathcal{S},\mathcal{S}^{\ast}) \nonumber\\
& &\!{}{\times}\,
\{\mathcal{C}_{\mathbf{w}}^{\ast}\,{+}\,\mathcal{C}_{\mathbf
{w}}(\mathcal{S},\mathcal{S}^{\ast})^{T}(\bolds{\Sigma}
_{\tilde{\bolds{\varepsilon}}}\,{+}\,\mathbf{I}_{n}\,{\otimes}\,
\bolds{\Sigma}_{\bolds{\varepsilon}})^{-1}\mathcal{C}_{\mathbf
{w}}(\mathcal{S},\mathcal{S}^{\ast}) \}^{-1}\\
&&\!{}{\times}\,\mathcal{C}_{\mathbf{w}}(\mathcal{S},\mathcal{S}^{\ast
})^{T}(\bolds{\Sigma}_{\tilde{\bolds{\varepsilon}}}\,{+}\,\mathbf{I}
_{n}\,{\otimes}\,\bolds{\Sigma}_{\bolds{\varepsilon}})^{-1}. \nonumber
\end{eqnarray}
The determinant is computed as
%
\begin{eqnarray}\label{eqdet}
&& \operatorname{det}(\bolds{\Sigma}_{\tilde{\mathbf{w}}}+\bolds
{\Sigma}_{\tilde{\bolds{\varepsilon}}}+\mathbf{I}_{n}\otimes
\bolds{\Sigma}_{\bolds{\varepsilon}})\nonumber \\
&&\qquad
=
\operatorname{det}\{\mathcal{C}_{\mathbf{w}}^{\ast}+\mathcal
{C}_{\mathbf{w}}(\mathcal{S},\mathcal{S}^{\ast
})^{T}(\bolds{\Sigma}_{\tilde{\bolds{\varepsilon}}}+\mathbf
{I}_{n}\otimes
\bolds{\Sigma}_{\bolds{\varepsilon}})^{-1}\mathcal{C}_{\mathbf
{w}}(\mathcal{S},\mathcal{S}^{\ast})\}\\
&&\qquad\quad{} \times
\{\operatorname{det} (\mathcal{C}_{\mathbf{w}}^{\ast})\}^{-1}
\operatorname{det}(\bolds{\Sigma}_{\tilde{\bolds{\varepsilon
}}}+\mathbf{I}_{n}\otimes\bolds{\Sigma}_{\bolds{\varepsilon}}).\nonumber
\end{eqnarray}
Notice that $\bolds{\Sigma}_{\tilde{\bolds{\varepsilon}}}+\mathbf
{I}_{n}\otimes\bolds{\Sigma}_{\bolds{\varepsilon}}$
is a sparse matrix and
$\mathcal{C}_{\mathbf{w}}^{\ast}+\mathcal{C}_{\mathbf{w}}(\mathcal
{S},\mathcal{S}^{\ast})^{T}(\bolds{\Sigma}
_{\tilde{\bolds{\varepsilon}}}
+\mathbf{I}_{n}\otimes\bolds{\Sigma}_{\bolds{\varepsilon
}})^{-1}\mathcal{C}_{\mathbf{w}}(\mathcal{S},\mathcal{S}^{\ast
})$ is an
$mR\times mR$ matrix. By letting $m$ be much smaller than~$n$
and letting $\bolds{\Sigma}_{\tilde{\bolds{\varepsilon}}}+\mathbf
{I}_{n}\otimes\bolds{\Sigma}_{\bolds{\varepsilon}}$
have a big proportion of zero entries, the matrix
inversion and the determinant in \eqref{eqswm-inv} and
\eqref{eqdet} can be efficiently computed. These computational
devices are combined with the techniques described in
subsection~\ref{ModelFit} for Bayesian model inference and spatial
prediction.

Now we consider one choice of the modulating function, that is, based on
a local partition of the domain.
Let $B_1, \ldots,B_{K}$ be $K$ disjoint subregions which divide the
spatial domain $D$. The modulating function is taken to be
$\mathcal{K}(\mathbf{s},\mathbf{s}')=\mathbf{1}_{K\times K}$
if~$\mathbf{s}$ and~$\mathbf{s}'$ belong to the same subregion,
and $\mathcal{K}(\mathbf{s},\mathbf{s}')=\mathbf{0}_{K\times K}$
otherwise.
Voronoi tessellation is one option to construct
the disjoint subregions [\citet{green1978computing}]. This
tessellation is defined by a number of centers $\mathbf{c}= (\mathbf
{c}_1,\ldots,
\mathbf{c}_{k})$, such that points within $B_i$ are closer to $\mathbf
{c}_i$ than any
other center $\mathbf{c}_j$, $j \neq i$,
that is, $B_i = \{\mathbf{s}: \|\mathbf{s}- \mathbf{c}_{i}\| \leq\|
\mathbf{s}- \mathbf{c}_{j}\|,
j\neq i\}$. Our choice of a Voronoi partitioning scheme is made on
the ground of tractability and computational simplicity.
We would like to point out that our methodology is not
restricted to this choice and any appropriate partitioning strategy
for the spatial domain could be adopted. Since the modulating
function so specified will generate an approximated covariance matrix
with block-diagonal structure, this version of the FSA method is referred
to as the FSA-Block.

The FSA-Block method provides an exact error correction for the predictive
process within each subregion, that is, $\Gamma_{\mathbf{w}}^{\dag
}(\mathbf{s}, \mathbf{s}')
=\Gamma_{\mathbf{w}}(\mathbf{s}, \mathbf{s}')$ if $\mathbf{s}$ and
$\mathbf{s}'$ belong to the same
subregion, and $\Gamma_{\mathbf{w}}^\dag(\mathbf{s}, \mathbf{s}')
=\Gamma_{\tilde{\mathbf{w}}}(\mathbf{s},
\mathbf{s}')$ if $\mathbf{s}$ and $\mathbf{s}'$ belong to different
subregions.
Unlike the
independent blocks analysis of spatial Gaussian process models, the
FSA can take into account large/global scale dependence across
different subregions due to the inclusion of
the predictive process component. Unlike the predictive process,
the FSA can take into account the small scale dependence due to the
inclusion of the residual process component.
An interesting special case of the FSA-Block is obtained by taking $n$
disjoint subregions, each of which contains only one observation.
In this case, $\tilde{\bolds{\varepsilon}}(\mathbf{s})$ is reduced to
an independent Gaussian
process, that is,
$\Gamma_{\tilde{\bolds{\varepsilon}}}(\mathbf{s},\mathbf{s})=\Gamma
_{\mathbf{w}}(\mathbf{s},\mathbf{s})-\Gamma_{\tilde{\mathbf
{w}}}(\mathbf{s}
,\mathbf{s})$,
and $\Gamma_{\tilde{\bolds{\varepsilon}}}(\mathbf{s},\mathbf{s}')=0$
for $\mathbf{s}\neq\mathbf{s}'$.
In fact, this special case corresponds to the modified predictive
process by
\citet{Finley2008}, that is, introduced to correct the bias for the
variance of the predictive process models at each location. Clearly,
our FSA-Block is a more general approach since it also corrects the
bias for
the cross-covariance between two locations that are located in the
same subregion.

For the FSA-Block, the computation in \eqref{eqswm-inv} and
\eqref{eqdet} can be further simplified.
Assuming the set of observed locations is grouped by subregions, that is,
$\mathcal{S}=\{\mathcal{S}_{B_{1}},\ldots,\mathcal{S}_{B_{K}}\}$, the
$nR\times nR$
matrix $\bolds{\Sigma}_{\tilde{\bolds{\varepsilon}}}$ becomes a
$K\times K$ block diagonal
matrix with the $k$th diagonal block
\[
\mathcal{C}_{\mathbf{w}}(\mathcal{S}_{B_{k}},\mathcal
{S}_{B_{k}})-\mathcal{C}_{\mathbf{w}}(\mathcal{S}_{B_{k}},\mathcal{S}
^{\ast})\mathcal{C}_{\mathbf{w}}^{\ast
-1}(\bolds{\theta})\mathcal{C}_{\mathbf{w}}^{T}(\mathcal
{S}_{B_{k}},\mathcal{S}^{\ast}),\qquad k=1,\ldots,K,
\]
and, thus,
$\bolds{\Sigma}_{\tilde{\bolds{\varepsilon}}}+\mathbf{I}_{n}\otimes
\bolds{\Sigma}_{\bolds{\varepsilon}}$ is a $K\times
K$ block diagonal matrix.
The inversion and determinant of this block diagonal matrix can be
directly computed efficiently without resorting to general-purpose
sparse matrix algorithms if the size of each block is not large.
%

The computational complexity of the FSA-Block depends on the knot
intensity and the block size. If we take equal-sized blocks,
then the computational complexity of the log likelihood calculation
is of the order $O(nm^2R^3+nb^2R^{3})$, where $m$ is the number of knots
and $b$ is the block size. This is much smaller than the original
complexity of $O(n^3R^3)$ without using covariance approximation.
Moreover, the computational complexity of the FSA
can be further reduced using parallel computation by taking advantage
of the block diagonal structure of $\bolds{\Sigma}_{\tilde{\bolds
{\varepsilon}}}$.

An alternative choice of the modulating function in
\eqref{eqtapering} is to use a positive definite function, that is,
identically zero whenever $\|\mathbf{s}-\mathbf{s}'\|\geq\gamma$.
Such a function is usually called a taper function and $\gamma$ is
called the taper range. The resulting FSA method is referred to as the
FSA-Taper.
In the univariate case, any compactly supported correlation function
can serve as a taper function, including the spherical and a family of
Wendland functions [see, e.g., \citet
{wendland1995piecewise};
\citet{wendland1998error};
\citet{gneiting2002compactly}].
\citet{sanghuang10} studied the FSA-Taper and demonstrated its usage
for univariate spatial processes.
For the multivariate processes considered in this paper,
the modulating function $\mathcal{K}(\mathbf{s},\mathbf{s}')$ can be
chosen as any valid multivariate taper function.
One such choice is the matrix direct sum of univariate
taper functions, that is, $\mathcal{K}(\mathbf{s},\mathbf{s}')=
{\bigoplus}_{r=1}^R K_{r}(\mathbf{s},\mathbf{s}';\gamma_{r})$, where
$K_{r}$ is a valid univariate taper function with the taper range
$\gamma_r$ used for the $r$th spatial variable, and different taper
ranges can be used for different variables.
This cross-independent taper function will work well with the FSA if
the cross dependence between co-located variables can be
mostly characterized by the reduced rank process $\tilde{\mathbf{w}}$.
Using this taper function, the cross-covariance matrix
of the residual process, $\bolds{\Sigma}_{\tilde{\bolds{\varepsilon
}}}$, can be transformed by
row and column permutations to a block-diagonal matrix with $R$
diagonal blocks, whose $r$th diagonal block is an $n\times n$ sparse
matrix with the $(i,j)$-entry being
$\operatorname{Cov}(\tilde\varepsilon_{r}(\mathbf{s}_i),\tilde
\varepsilon_{r}(\mathbf{s}_j))
= \{\operatorname{Cov}(w_{r}(\mathbf{s}_i),w_{r}(\mathbf{s}_{j}))-
\operatorname{Cov}(\tilde{w}_{r}(\mathbf{s}_i),\tilde
{w}_{r}(\mathbf{s}_{j}))\}
K_{r}(\mathbf{s}_i,\mathbf{s}_j;\gamma_{r})$,
where $\tilde{w}_{r}(\mathbf{s})$ is the reduced rank predictive
process for the
$r$th spatial variable and $\tilde\varepsilon_{r}(\mathbf{s})$ is the residual
process for the $r$th spatial variable.
If we take the same taper range for each spatial variable, the computational
complexity of the log likelihood calculation is of the order
$O(nm^2R^3+ng^2R)$, where $g$ is the average number of nonzero entries per
row in the $n\times n$ residual covariance matrix
$[\operatorname{Cov}(\tilde\varepsilon_{r}(\mathbf{s}_i),\tilde
\varepsilon_{r}(\mathbf{s}_j))]_{i,j=1}^{n}$
for the $r$th spatial variable. This is a substantial reduction from the
original computational complexity of $O(n^3R^3)$ without using the
covariance approximation.

Use of the FSA involves the selection of knots and the local
partitioning or tapering strategy. Given the number of knots, we
follow the suggestions by \citet{banerjee2010hierarchical} to use the
centers obtained from the $K$-means clustering as the knots
[e.g., \citet{kaufman1990finding}].
A careful treatment of the choice of knot intensity $m$ and
the number of partitions $K$ or the taper range $\gamma$
will offer good approximation to the original covariance function.
Apparently, a denser knot intensity
and larger block size or larger taper range will lead to better
approximation, at higher computational cost. In principle, we will
have to implement the analysis over different choices of $m$ and $K$
or $\gamma$
to weigh the trade-off between inference accuracy and computational cost.
We have used the Euclidean
distance and taken~$m$ to be 225, $K$ to be 36, and $\gamma$ to be 10
for the spherical taper function in our simulation
study, and used the chordal distance and taken $m$ to be 200 and~$K$ to be 10 in the real application and have found that such choices
work well. A full discussion of this issue will be undertaken in future
work.

\section{Simulation results}\label{Sim}

In this section we report results from a simulation study to
illustrate the performance of the FSA approach and compare it with
the predictive process approach and the independent blocks analysis.
The computer implementation of all the
approaches used in this simulation study and the analysis of multiple
climate models in the following section were written in Matlab and
run on a processor with dual 2.8 GHz Xeon CPUs and 12~GB memory.

In this simulation study we generated $n = 2\mbox{,}000$ spatial locations
at random from the square $[0,100]\times[0,100]$.
The data generating model is a bivariate LMC model with a constant
$2\times2$ lower triangular transformation matrix $\mathbf{A}$,
\begin{equation}
\mathbf{Y}(\mathbf{s}) = \mathbf{A}\mathbf{U}(\mathbf{s}) + \bolds
{\varepsilon}(\mathbf{s}),\nonumber
\end{equation}
where $\bolds{\varepsilon}(\mathbf{s}) \sim\operatorname{MVN}(\mathbf
{0},\tau^{2}\mathbf{I}_{2})$,
and $\mathbf{U}(\mathbf{s})=[U_{q}(\mathbf{s})]_{q=1,2}$ is a
$2\times1$ process with two
independent components, each of which is a univariate spatial
process with mean zero, unit variance and exponential correlation
function. The range parameters for $U_{1}(\mathbf{s})$ and
$U_{2}(\mathbf{s})$ are
$\phi_{1}=10$ (i.e., such that the spatial correlation is
about $0.05$ at 30 distance units) and $\phi_{2}=20$, respectively.
The diagonal elements of $\mathbf{A}$ are $a_{11}=1$ and $a_{22}=0.5$, and
the nonzero off-diagonal element of $\mathbf{A}$ is $a_{21}=0.5$.
We set the nugget variance $\tau^2=0.01$.

Given these data, we used the Bayesian MCMC approach to generate
samples from the posterior distributions of the model parameters.
We assigned $\operatorname{Unif}(1, d_{\mathrm{max}}/3)$ priors to
the range
parameters $\phi_{1}$ and $\phi_{2}$, where $d_{\mathrm{max}}$ is the
maximum distance
of all pairs. The diagonal elements of $\mathbf{A}$ and $\tau^2$ were assumed
to have the inverse gamma distribution with the shape parameter~2 and
the scale parameter 1, $\operatorname{IG}(2,1)$, as priors. We
assigned a normal
prior with large variance for~$a_{21}$.

%
\begin{table}
\tabcolsep=3pt
\caption{The mean and the standard deviations (in parentheses) of
the model parameters for the full covariance
model, the predictive process approximation, the independent blocks
approximation, the FSA-Block and the FSA-Taper}\label{BayesCompare}
\fontsize{7.5}{10.5}{\selectfont
\begin{tabular*}{\textwidth}{@{\extracolsep{\fill}}lcccccc@{}}
\hline
\textbf{Model} & $\bolds{\phi_1}$ & $\bolds{\phi_2}$ & $\bolds{a_{11}}$ & $\bolds{a_{12}}$ & $\bolds{a_{22}}$ & $\bolds{\tau^2}$
\\
\hline
True & 10\phantom{0000000.} & 20\phantom{0000000.} & 1\phantom{0000000.} & 0.5\phantom{000000.} & 0.5\phantom{000000.} & 1.00e--2\phantom{00000000.} \\
Full model & 10.47 (1.02) & 22.07 (4.69) & 0.99 (0.05) &0.51 (0.03)&0.52
(0.05)& 1.01e--2 (3.90e--4) \\
Predictive process & 13.32 (2.32)&22.71 (7.35)&1.22 (0.08)&0.66 (0.05)&0.49 (0.06)&0.15 (3.46e--3)\phantom{000}\\
\quad$m=225$& & & & & & \\
Independent blocks& \phantom{0}4.47 (0.99)&\phantom{0}4.94 (1.16)&3.56 (0.39)&1.44 (0.28)&1.96 (0.14)&8.97e--3 (1.18e--3)\\
\quad$k=36$& & & & & & \\
FSA-Block & 11.36 (2.21)&21.17 (5.24)&1.02 (0.09)&0.53 (0.05)&0.49 (0.05)&1.10e--2 (9.00e--4)\\
\quad$m=225,k=36$& & & & & & \\
FSA-Taper& 14.89 (1.90)&29.92 (7.41)&1.05 (0.07)&0.58 (0.04)&0.50 (0.05)&8.36e--3 (1.07e--3)\\
\quad$m=225,\gamma=10$& & & & & & \\
\hline
\end{tabular*}}
\end{table}

We compared the model fitting results from four covariance
approximation methods: the predictive process, the independent
blocks approximation, the FSA-Block and the FSA-Taper.
As a benchmark for our comparison, we also fit the original
full covariance model without using a covariance
approximation. For the predictive process approximation, we
used $m=225$ knots. For the independent blocks approximation, we
used $K=36$ blocks.
For the FSA-Block, we used $m=225$ knots and $K=36$ blocks.
For the FSA-Taper, we used $m=225$ and a spherical tapering function
with taper range $\gamma=10$. The number of blocks for the FSA-Block
and the taper range for the FSA-Taper are selected such that these two
FSA methods lead to comparable approximations for the small scale
residual covariance. For the above methods, the knots were chosen
as the centers from the $K$-means clustering, and the blocks were taken
as equal-sized squares that form a partition of
the $[0,100]\times[0,100]$ region.

Table \ref{BayesCompare} displays the Bayesian posterior means and the
corresponding posterior standard deviations for the model parameters under
each approach. We observe that the diagonal values of $\mathbf{A}$,
the range
parameter of $U_1$, and the nugget variance $\tau^2$ are
all overestimated by the predictive process. We also notice large
biases of the parameter estimates using independent blocks approximation.
The FSA-Block provides the most accurate parameter estimation
among all the methods.

%
\begin{table}
\tabcolsep=3pt
\caption{DIC scores and MSPEs for the full covariance
model, the predictive process approximation, the independent blocks
approximation, the FSA-Block and the FSA-Taper. MSPE-Random is based on
a test data set of 200 locations
that are randomly selected from $[0,100]\times[0,100]$. MSPE-Hole is
based on a test data set of size 200 that consists of 160 randomly
selected locations from $[0,100]\times[0,100]$ and 40 random
locations within two circles:
$\{(x,y);x^2+(y-90)^2<30\}$ and $\{(x,y);(x-50)^2+(y-50)^2<35\}$}\label{BayesCompare2}
\fontsize{8}{10.5}{\selectfont
\begin{tabular*}{\textwidth}{@{\extracolsep{4in minus 4in}}lccccc@{}}
%
\hline
&  & \multicolumn{1}{c}{\textbf{Predictive}} & \multicolumn{1}{c}{\textbf{Independent}} & &
\\
&  & \multicolumn{1}{c}{\textbf{process}} & \multicolumn{1}{c@{}}{\textbf{blocks}} & \multicolumn
{1}{c}{\textbf{FSA-Block}}  & \multicolumn{1}{c@{}}{\textbf{FSA-Taper}}\\[-6pt]
& & \multicolumn{1}{c}{\hrulefill} & \multicolumn{1}{c}{\hrulefill} & \multicolumn
{1}{c}{\hrulefill} & \multicolumn{1}{c@{}}{\hrulefill} \\
&\textbf{Full model} & $\bolds{m=225}$ & $\bolds{K=36}$ & $\bolds{m=225}$, $\bolds{K=36}$ & \multicolumn{1}{c@{}}{$\bolds{m=225}$, $\bolds{\gamma=10}$} \\
\hline
DIC & 871 & 2357 & 3791 & 918 & 1547\\
MSPE-Random & 0.12 & 0.17 & 0.14 & 0.12 & 0.12\\
MSPE-Hole & 0.16 & 0.26 & 0.26 & 0.18 & 0.18\\
\hline
\end{tabular*}}
\end{table}

To gauge the performance on model fitting for different approaches,
we used the deviance information criterion
[DIC, \citet{spiegelhalter2002bmm}], which is easily calculated from
the posterior samples. From Table~\ref{BayesCompare2}, we observe
that the benchmark full covariance model has the smallest DIC score,
indicating the best model fitting. The FSA-Block approach gives a~slightly larger DIC score than the full model, while the predictive
process and the independent blocks approximation yield
significantly larger DIC scores than the FSA and the full model.
This result shows that the FSA
performs much better than the predictive process and independent
blocks approximation in terms of model fitting.

To compare the methods in terms of prediction performance,
we computed the mean square prediction errors (MSPE) based on
a simulated test data set of 200 locations using the previously
simulated data set with observations at 2000 locations as the
training set. We experimented with two different kinds of test sets:
one set consists of 200 locations randomly selected from
$[0,100]\times[0,100]$ and another consists of 160 randomly selected
locations from $[0,100]\times[0,100]$ and 40 random locations within
two circles: $\{(x,y);x^2+(y-90)^2<30\}$ and
$\{(x,y);(x-50)^2+(y-50)^2<35\}$. The second interpolation
scenario is common in practice where missing data often correspond
to sizable gaps/holes.

From Table~\ref{BayesCompare}, we see that both the FSA-Block and
the FSA-Taper methods lead to much more accurate predictions than
the predictive process and the independent blocks approximation. In
the scenario of predicting for missing gaps/holes, the advantage of
using the FSA approach over the other two approximation approaches
is more significant.

\begin{figure}

\includegraphics{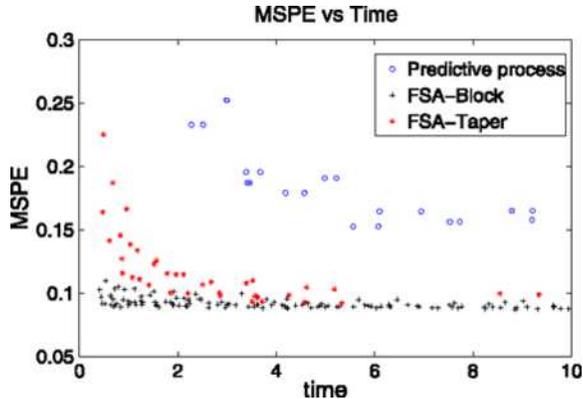}

\caption{The MSPE versus time plot for the simulation
study in Section~\protect\ref{Sim} under the predictive process (circle), the
FSA-Block (plus) and the FSA-Taper (star).}\label{mspetime}
\end{figure}

To compare the computation efficiency of the covariance approximations
for larger data sets, we repeated the simulation study when the
number of spatial locations in the training set is increased to 5,000
and the number of locations in the test set to 1,000. We measured
the MSPE and the associated computational time based on the
prediction at the 1,000 test locations. The test set consists
of 800 randomly sampled locations from $[0,100]\times[0,100]$ and 200
locations randomly selected
within two circles: $\{(x,y);x^2+(y-90)^2<30\}$ and
$\{(x,y);(x-50)^2+(y-50)^2<35\}$. The MSPE for each model was obtained by
plugging in the true parameter values into the BLUP equation.
Pairs of the MSPE and the computational time were obtained by
varying the knot intensity for the predictive process model, block size
for the independent blocks approximation, knot intensity and block size for
the FSA-Block approach, and knot intensity and taper
range for the FSA-Taper approach.
Figure~\ref{mspetime} shows that both the
FSA-Taper and the FSA-Block methods outperform the predictive process
in terms of their computational efficiency for making predictions. The
independent blocks approximation yields
much larger MSPE than the other three approaches given the same
computational time and we decided not to include its results in
Figure~\ref{mspetime}. It is also noticeable that the required
computational time of the FSA-Block approach to obtain the same MSPE
is much less than that of the FSA-Taper approach. For this reason,
we used the FSA-Block
approach to analyze the climate errors data in Section~\ref{secappresult}.

%
\begin{figure}

\includegraphics{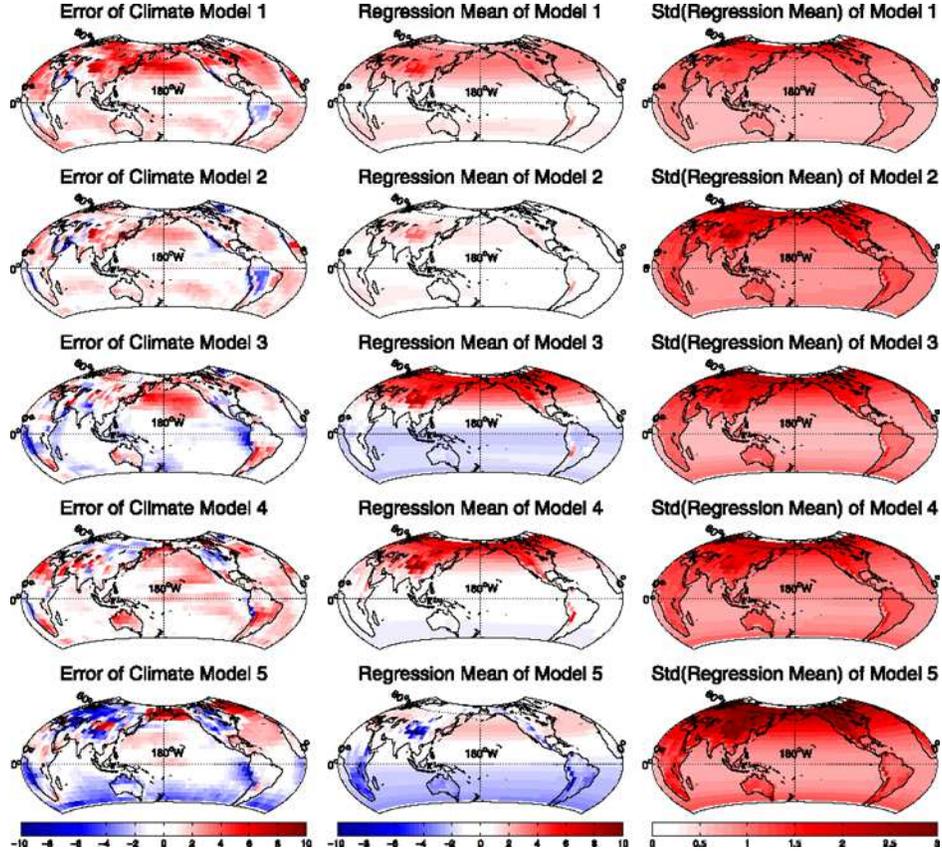}

\caption{The first column shows the surface maps
of climate model errors. The second and the third columns show the
estimated mean structure of each climate model error and the associated
standard deviations.}\label{estimatedmean}
\end{figure}

We acknowledge that performance for the approximation approaches
may depend on the characteristics of spatial data, such as sample
size, pattern of sampling locations, spatial smoothness and
spatial correlation range.
To investigate the effect of spatial correlation range on
the performance of the FSA approach, we conducted two other
experiments: one with small range parameters $\phi_1=5$ and $\phi
_2=10$, and the other with large range parameters $\phi_1=30$ and
$\phi
_2=60$. In the case of small spatial ranges, the independent blocks
approximation performed better than the predictive process, while the
FSA performed similarly to the independent blocks analysis.
In the case of large spatial ranges, the predictive process performed
similarly to the FSA and both methods performed significantly better than
the independent blocks approximation. These two experiments
indicate that both the predictive process and the independent blocks
approximation have
their modes of successes and failures. Under their failure modes, to achieve
accurate model inference and prediction, one has to use either a significantly
large rank number $m$ for the predictive process or a very small
number of blocks $K$ for the independent blocks approach,
therefore, the computational
advantages associated with a small $m$ and a large~$K$ would disappear.
In contrast, by the combining of a small $m$ and a large~$K\!$, the
FSA-Block
flexibly accommodates data sets with either large scale or small
scale spatial variations while still maintaining the computational efficiency.

\section{Application result}
\label{secappresult}

To build a joint model that accounts for the cross-covariance
structure of the multiple climate model errors, we fit and compared
two versions of the LMC models, corresponding to spatially fixed
and spatially varying transformation matrices, as described in
Section~\ref{secmodel}. The LMC model specification depends
on the ordering of the response variables because of the lower
triangular specification of the transformation matrix. When applying
the LMC model to the five climate model errors, we tried a few
orderings of climate models and found that different orderings
may produce different values of parameters but they produced fairly
consistent estimates of variances and cross-correlations. Therefore,
we chose to present the results based on the order given in
Table~\ref{list}.

We applied the FSA-Block approach to facilitate the computation.
We used 225 knots selected by the $K$-means clustering algorithm and
divided the study region into 10 regions for subsequent analysis.
We assumed the exponential spatial correlation function for each
$U_{q}(\mathbf{s})$, and assigned a uniform prior $U(50, 4\mbox{,}500)$ for
each of
the spatial range parameters given that the maximum chordal distance
between any two locations is 12\mbox{,}757~km. For each of the coefficients
$\bolds{\beta}$ in the regression in model (\ref{MultiSpatialRegressionEqn}),
we assumed independent normal priors with mean 0 and variance 1,000. For
the LMC model with a constant $\mathbf{A}$, we assumed the diagonal
entries to
have truncated normal distributions ranging from 0 to $\infty$ and
diagonal entries to have normal distributions, with their means being
the empirical estimates of $\mathbf{A}$ and variances 1,000. For the spatially
varying LMC model with $\mathbf{A}(\mathbf{s})$, we assumed that the
intercepts of
the coefficients $\bolds{\eta}_{ij}$ in the regression model for
$\mathbf{A}(\mathbf{s})$
have normal distributions, with their means being the empirical
estimates of $\mathbf{A}$ and variances being 1,000. We set normal
priors with
mean 0 and variance 1,000 to the other coefficients of $\bolds{\eta}_{ij}$.
For each model, we ran 3,000 iterations of MCMC to collect posterior
samples after a burn-in period of 1,000 iterations, thinning every third
iteration.

The DIC scores of two specifications for $\mathbf{A}$ were compared, one
with a~constant~$\mathbf{A}$ and the other with spatially varying
$\mathbf{A}$
depending on polynomials of latitude, land/ocean effect and altitude,
as detailed in Section~\ref{secmodel}.
The DIC score for the model with a spatially varying $\mathbf
{A}(\mathbf{s})$ is
9,901, which is much smaller than the score of 11,975 for the model with
a constant
$\mathbf{A}$. This suggests that the spatially varying LMC model performs
significantly better than the model with a constant~$\mathbf{A}$, as we
expected. From now on, we present results based solely on the
spatially varying LMC model.


Figure~\ref{estimatedmean} shows the estimated mean structure and its
standard deviations obtained from the posterior samples. It is
interesting to note a clear distinction between model 5 and the rest
of the models;
for model 5, the estimated means are negative with large magnitudes over
high altitude areas, whereas the rest gives large positive values for
the estimated mean structure over high latitude areas. This finding is
consistent with the result in \citet{junetal08} [note that the model
numbers in this paper and the model numbers in \citet{junetal08} are
different, although all the models used in this paper are also used in
\citet{junetal08}]. Recall that the model error is calculated by
subtracting the model output from the observation. The above result
suggests that models~1--4 may underestimate
the mean state of the surface temperature over the high-altitude
and high-latitude regions, and model 5 may overestimate the mean
state over the high-altitude area.
The estimated mean structure and its associated standard deviations are
quite similar for the models developed by the same groups---models 1
and 2 from the GFDL group of NOAA, and models 3 and 4 from the Hadley
Centre in the UK. The spatial patterns of the associated standard
deviations for those pairs are also quite similar. For all the models, altitude
is responsible for the dominant effects in the fixed part of the
process and we also see some effects of latitude, although longitude
does not seem to be significant in the mean structure. Models 3--5 seem
to have large errors in the high-latitude and sea-ice area. The
indicator for the land and the ocean is not significant for the mean
structure. This may be due to the fact that we already include
altitude as a covariate (altitude is positive over the land
and zero over the ocean).


\begin{figure}

\includegraphics{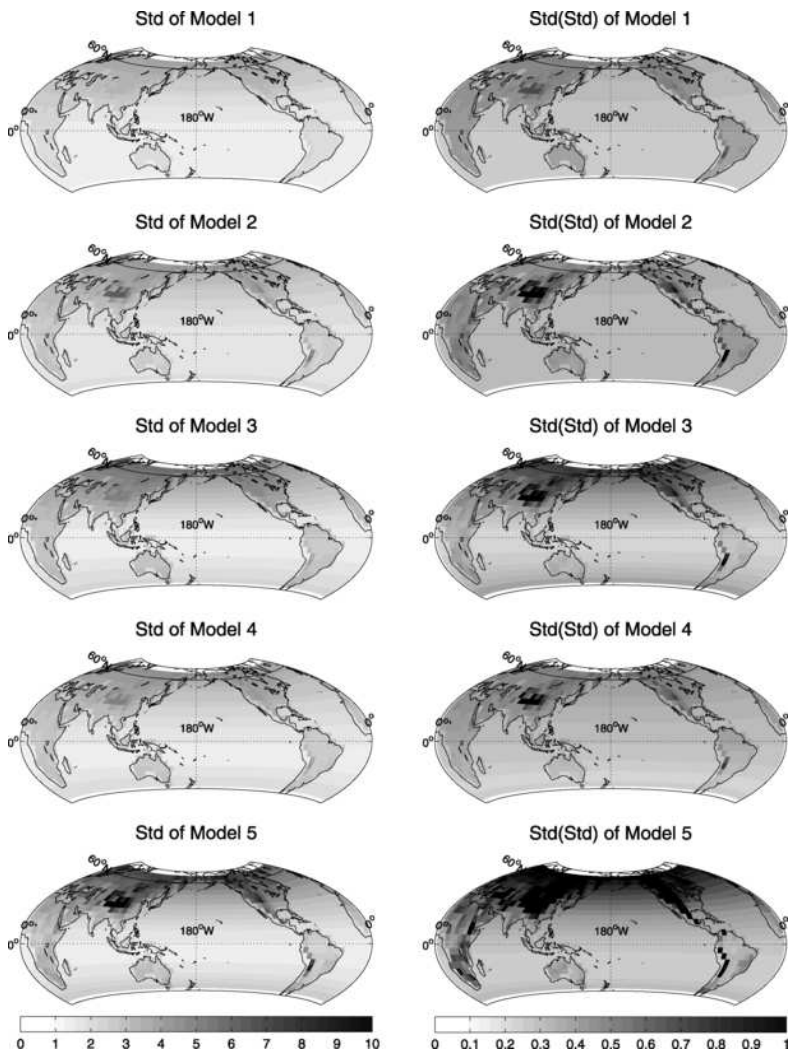}

\caption{Surface maps of the estimated standard
deviations of each climate model (left column) and the associated
standard deviations (right column).}\label{RealVarMap}
\end{figure}

Given the posterior samples of the model parameters, we draw samples
of the cross-covariance matrix at each location $\mathbf{s}$ based on
$\Gamma_{\mathbf{w}}(\mathbf{s},\mathbf{s})=\mathbf{A}(\mathbf
{s})\mathbf{A}^{T}(\mathbf{s})$. The diagonals of
$\Gamma_{\mathbf{w}}(\mathbf{s},\mathbf{s})$ are the variances of
the climate model errors at
location $\mathbf{s}$. Figure~\ref{RealVarMap} shows the maps of standard
deviations for each
model error and the standard deviations of the standard deviations obtained
from the posterior samples.
The patterns throughout all 5 models are quite consistent; high
standard deviations at high altitudes, and standard deviations are
higher over the land
than over the ocean. For models 1 and 3, latitude seems to be a
significant factor. For all the models, given the values of the
posterior sample standard deviations of the standard deviations, the
spatial patterns that we observe (such as high variances over high
altitude or high latitude area) are statistically significant and are
not due to random variations in the
data. The sea-ice region is one of the places that models in general
have trouble. Although we do not have a factor for sea-ice region in
the model, we see high standard deviations around sea-ice area. This
may be due
to the interaction between latitude and the indicator for the land and
the ocean. As we expected, model 5 has significantly larger
values of standard deviations, especially in high-altitude areas,
compared to the
rest of the models. The standard deviations of the error of model 5
over the Himalayan area are more than 10. Their associated posterior
sample standard
deviations are also quite large. Overall, we observe that standard
deviations of
model errors are slightly higher over the land than over the ocean.

%
\begin{figure}

\includegraphics{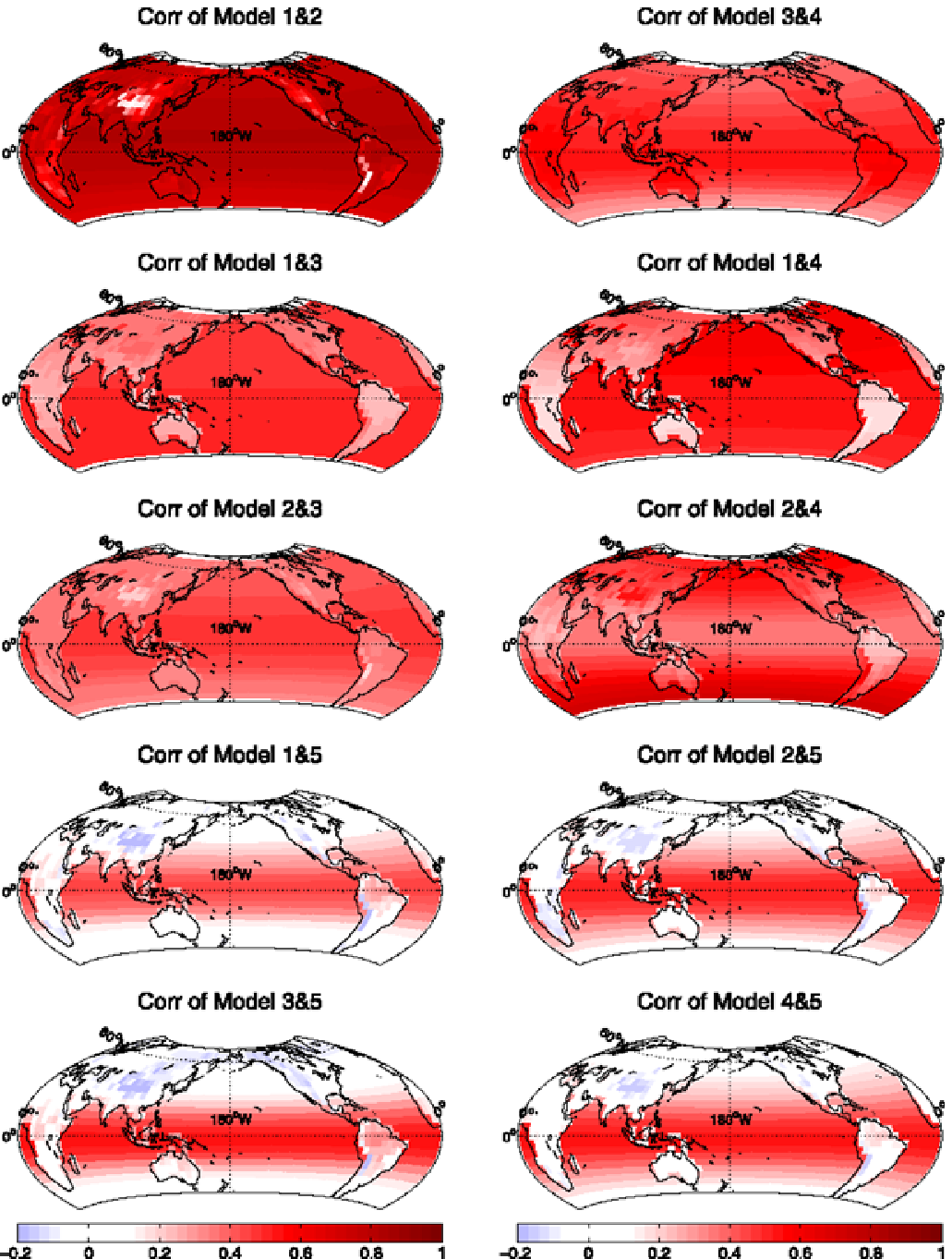}

\caption{Surface maps of the estimated
cross-correlation for each pair of climate model errors.}\label{estimatedcor}
\end{figure}

%
\begin{figure}

\includegraphics{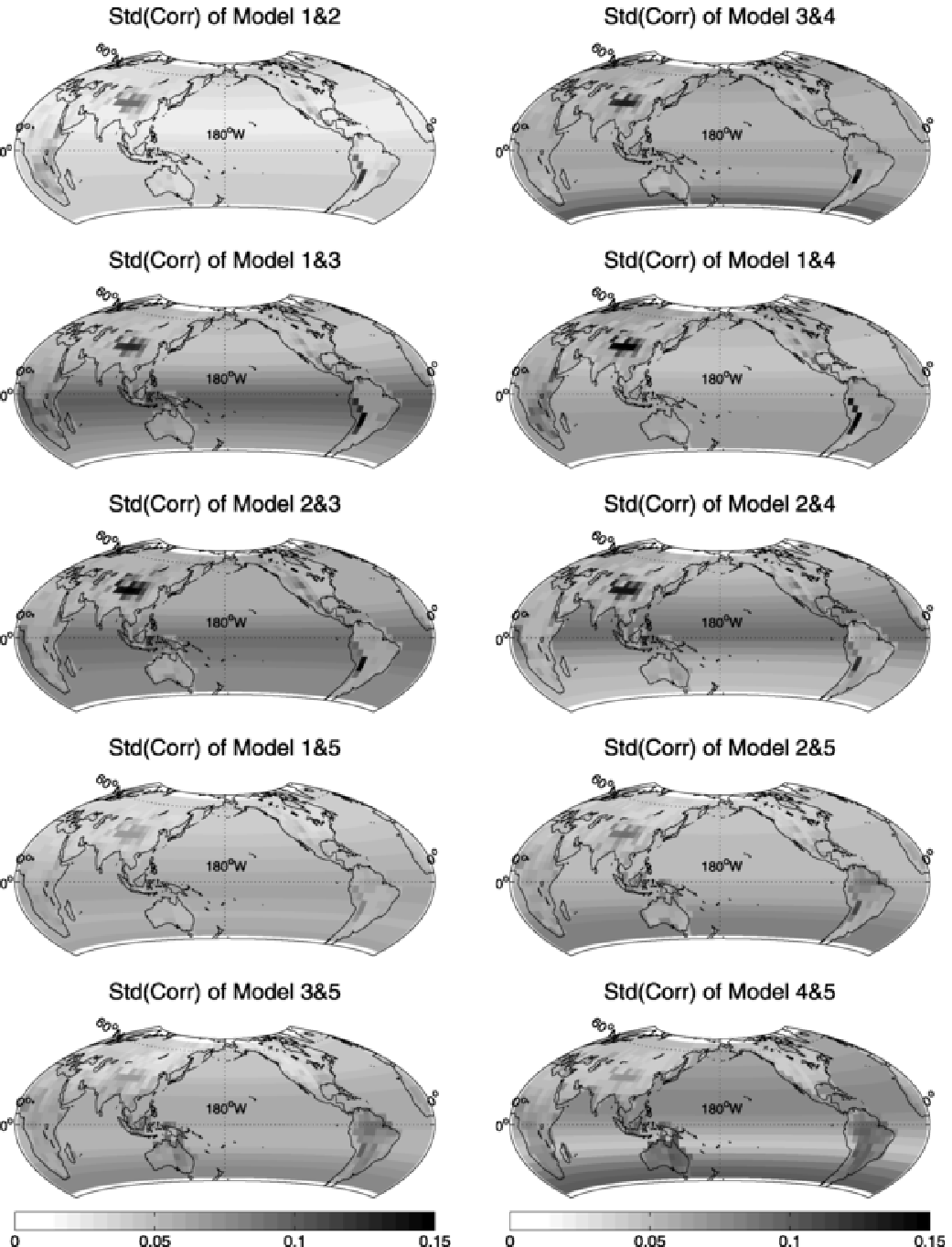}

\caption{Standard deviation of the estimated
cross-correlation for each pair of climate model errors in
Figure~\protect\ref
{estimatedcor}.}\label{stdcor}
\end{figure}

Now we discuss the estimates of cross-correlations of different
climate model errors. Figure~\ref{estimatedcor} gives the spatial maps
of cross-correlations between each pair of climate model errors,
obtained from the posterior samples.
The corresponding standard deviation of the estimated
cross-correlations using the posterior samples are given in
Figure~\ref{stdcor}. In both figures, we use the same color scale
across pairs of models to make the comparison easier. First, notice
that overall the correlation between models 1 and 2, two models
developed by the GFDL group of NOAA, is the highest. For models 1 and
2, overall the correlation values are strikingly high and the
associated standard
deviation values are close to zero. The maximum correlation value over
the entire domain is 0.769, the average of correlations over the ocean
is 0.732 with the standard deviation 0.017, and the average over the
land is 0.639 with the standard deviation 0.021. \citet{junetal08}
also report the highest level of cross-correlation for this pair of
models. The cross-correlations between models 3 and 4 (two models
developed by the Hadley Centre in the UK) are not as high as those between
models 1 and~2 and they are comparable to the cross-correlations
between any one of the models 1 and 2 and any one of the models 3 and 4. In
addition, the cross-correlation structure between models 1 and 2 shows
different spatial patterns compared with that between models 3
and 4. Models 1 and 2 have quite small correlation over high-altitude
area, meaning that the two models disagree over high-altitude areas,
whereas models~3 and~4 have consistently large correlations over the
entire domain. For models~3 and~4, the average of correlations over
the land and the ocean are both slightly larger than~0.4.
Cross-correlations between model 5 and the rest of the models are
quite small in magnitude and the patterns are consistent for all the 4
pairs. Correlations between model 5 and the rest of the models are
quite small over the land, over high latitudes in the Northern
Hemisphere and over low latitude area in the Southern Hemisphere. From
these maps,
it is clear that model 5 is significantly different from the rest of
the models and this agrees with the result in~\citet{junetal08}. In
many pairs of models, we see clear effects of the indicator for the land
and the ocean, and the latitude.

\begin{figure}

\includegraphics{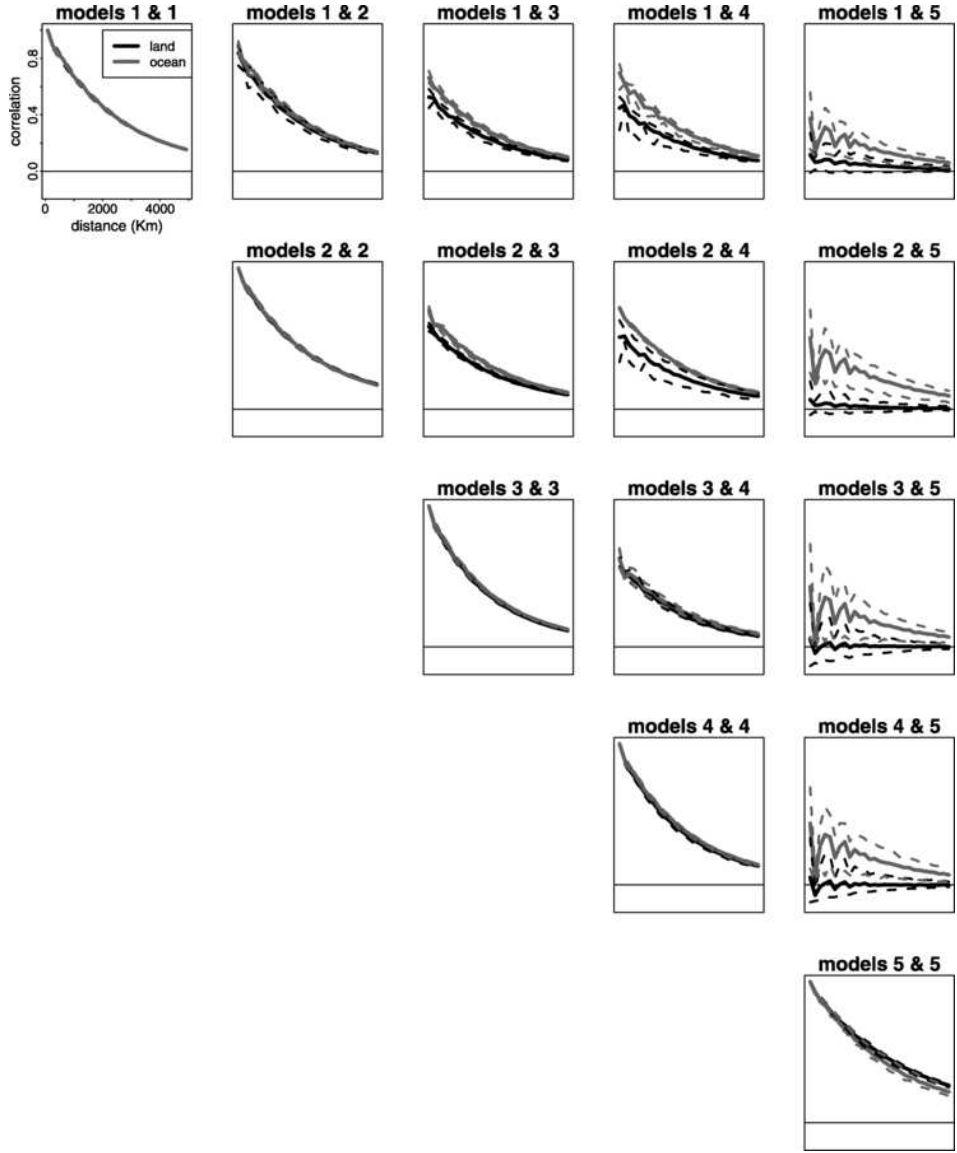}

\caption{Spatial correlations for pairs of model
errors against chordal distance (unit: km). Due to the nonstationarity
of the correlation structure, we display the correlation in the
following way. For each pair of model errors, we first obtain a
$1\mbox{,}656\times1\mbox{,}656$ cross (or marginal)-correlation matrix given from the
posterior mean of the correlations. Then we calculate the averages, the
10th and the 90th percentiles of the correlations within each bin, with
30 equally spaced bins from distance 0 to 5,000~km, separately over the
land and the ocean. Solid lines connect the binned averages and dashed
lines connect the 10th and the 90th percentiles.}\label{corlandsea}
\end{figure}

In addition to the cross-correlations at the same location, our method
allows us to examine both marginal and cross-covariances/correlations
between two different locations, based on the posterior samples. Given
that the cross-correlations at the same location are quite different
over the land and the ocean, we examined the spatial correlations
against spatial distance over the land and the ocean separately.
Figure~\ref{corlandsea} gives both marginal and cross-correlations
against chordal distance (unit: km) for pairs of models based on the
1,656 observational grid points. Since we adopted exponential spatial
correlation functions for the latent spatial processes $U_{q}(\mathbf{s})$'s
in the LMC model with the lower triangular structure for $\mathbf
{A}(\mathbf{s})$,
the fitted correlation function for model 1 is isotropic. We also note
that marginally the model errors 2, 3 and 4 give similar spatial
correlation structures over the land and the ocean, while model 5 shows
mild discrepancy between
the correlations over the land and the ocean. Moreover, the spatial
correlation structure of model 5 exhibits a slower spatial decay
compared with the rest of the climate model errors. Models developed by
the same group (i.e., pairs 1 and~2, and 3 and 4) give similar
spatial cross-correlation structure for both over the land and over the
ocean. The cross-correlations between model 5 and the rest of the
models are close to zero over the land, while those over the ocean are
significantly different from zero.

As in \citet{junetal08} and \citet{sainetal10}, we can perform the
analysis for the summer season average of Northern Hemisphere
(JJA, June--July--August average) in the same way that we did for the DJF
averages. We do not include those results for brevity of the paper.

%
\begin{figure}

\includegraphics{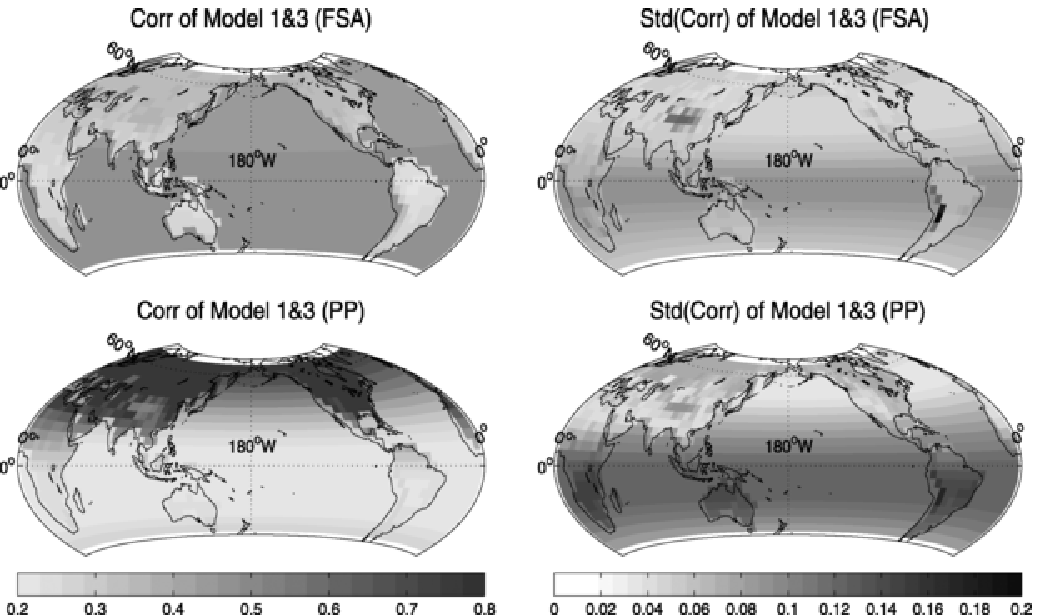}

\caption{Correlation maps and the associated standard
deviation maps between models 1 and 3. The upper panel shows the maps
using the FSA-Block approach and the bottom panel shows the maps using
the predictive process.}\label{Corr1n3}
\end{figure}

We have also implemented the spatially varying LMC using the
predictive process with the same knots as the FSA ($m=200$)
for comparison purposes. The DIC score of the predictive process model
is 35,526, which is much larger than the DIC score of 9,901 of the FSA
model, indicating that the FSA model has a better model fitting for
the data. We did not observe significant difference in the estimations
for the spatial range parameters. We present in Figure~\ref{Corr1n3}
the correlation maps between models 1 and 3 to illustrate the
difference between the results of the two approaches.
Although, for this real data analysis, the true
correlations between models 1 and 3 are unknown, the results obtained
from the FSA approach seem to be more reasonable than those from the
predictive process. For instance, the map using the FSA approach
clearly shows that the correlations over the ocean are in general
higher than the correlations over the land, as expected based on
previous studies. The map using the predictive process
fails to show this pattern. The poor performance of the
predictive process in this example might be due to the inadequate
number of knots and hence can be remedied by increasing the knots
intensity. However, increasing the number of knots will greatly increase
the computational time of the predictive process.

\section{Discussion}\label{sec6}

We built a joint spatial model for multivariate climate model errors
accounting for their cross-covariances through a spatially varying LMC
model. To facilitate Bayesian computation for large spatial data sets,
we developed a covariance approximation method for multivariate
spatial processes. This full-scale approximation can capture both the
large scale and small scale spatial dependence and correct the bias
problem of the predictive process.

Our empirical results confirmed that pairs of climate models developed
by the same group have high correlations and climate models in general
have correlated errors. We also showed that some climate models are very
different from other climate models and, thus, the cross-correlations
between them are quite small.

In principle, we could combine multiple climate model outputs as a
weigh\-ted linear combination, along the same lines as \citet
{sainfurrer10}, based on
our modeling approach rather than the Bayesian Gaussian MRF model
described in Section 4 of \citet{sainfurrer10}.
However, recently there have been
several papers that raise concerns about the practice of combining multiple
climate model outputs through model weighting or ranking [see, e.g.,
\citet{Knutetal10};
\citet{knutti10};
\citet{weigeletal10}]. The main concern is the
difficulty of interpreting the weighted average of climate models physically.
\citet{goodpractice} suggest that if model ranking or weighting is
applied, both the quality metric and the statistical framework used to
construct the ranking or weighting should be recognized. To determine
what is the best quality metric in weighting or ranking the
climate models is a challenging problem.

We focus on the climate model errors in this paper. It would be
interesting to build more elaborate joint models of multiple climate
model outputs as well as observations at their original spatial grid
resolutions. To follow this path, we need to have a statistical
representation of the true climate, which is very challenging.
One possibility is to model the observation as the truth, with
measurement errors assumed with a simple covariance structure.
However, this assumption might not be realistic considering
the relatively complex nature of biases and errors in the climate
observations.

In this paper we addressed only the spatial aspect of the problem and we
applied our methodology to the mean state of the climate variable. It
would be interesting to extend our approach to spatio-temporal
problems and, in particular, to consider the climate model outputs
in their original time scale of monthly averages or annual averages
for studying the trend of the climate model errors.

\section*{Acknowledgments}
The authors thank the Editor, the Associate
Editor and two referees for valuable suggestions that improved the paper.

%

%

%


\printaddresses

\end{document}